%% file: paper.tex
\documentclass[]{jfm}

\usepackage{graphicx}
\usepackage{booktabs}
\usepackage{placeins}
\usepackage{xcolor}
\usepackage{subcaption}
\usepackage{newtxtext}
\usepackage{newtxmath}
\usepackage{amsmath}
\usepackage{natbib}
\usepackage{hyperref}
\hypersetup{
    colorlinks = true,
    urlcolor   = blue,
    citecolor  = black,
}

\newcommand{\RomanNumeralCaps}[1]
\linenumbers

\usepackage{tikz,xcolor,hyperref}

\definecolor{lime}{HTML}{A6CE39}
\DeclareRobustCommand{\orcidicon}{
	\begin{tikzpicture}
	\draw[lime, fill=lime] (0,0) 
	circle [radius=0.16] 
	node[white] {{\fontfamily{qag}\selectfont \tiny ID}};
	\draw[white, fill=white] (-0.0625,0.095) 
	circle [radius=0.007];
	\end{tikzpicture}
	\hspace{-2mm}
}
	
\foreach \x in {A, ..., Z}{\expandafter\xdef\csname orcid\x\endcsname{\noexpand\href{https://orcid.org/\csname orcidauthor\x\endcsname}
			{\noexpand\orcidicon}}
}

\title{Edge-effects in the turbulent flow over flexible aquatic vegetation}

\author{Giulio Foggi Rota\aff{1}, Elisa Tressoldi\aff{1,2}, Francesco Avallone\aff{2}, and Marco Edoardo Rosti\aff{1,}\textsuperscript{\corresp{\email{marco.rosti@oist.jp}}}}

\affiliation{
\aff{1} Complex Fluids and Flows Unit, Okinawa Institute of Science and Technology Graduate University (OIST), 1919-1 Tancha, Onna-son, Okinawa 904-0495, Japan
\aff{2} Dipartimento di Ingegneria Meccanica e Aerospaziale, Politecnico di Torino, Corso Duca degli Abruzzi 24, 10129, Torino, Italy
}

\begin{document}
\maketitle

\begin{abstract}
Riparian vegetation along riverbanks and seagrass along coastlines interact with water currents, significantly altering their flow. To characterise the turbulent fluid motion along the streamwise--edge of a region covered by submerged vegetation (\textit{canopy}), we perform direct numerical simulations of a half--channel partially obstructed by flexible stems, vertically clamped to the bottom wall. An intense streamwise vortex forms along the canopy edge, drawing high--momentum fluid into the side of the canopy and ejecting low--momentum fluid from the canopy tip, in an upwelling close to the canopy edge. This mechanism has a profound impact on the mean--flow and on the exchange of momentum between the fluid and the structure, which we thoroughly characterise. The signature of the canopy--edge vortex is also found in the dynamical response of the stems, assessed for two different values of their flexibility. Varying the flexibility of the stems, we observe how different turbulent structures over the canopy are affected, while the canopy--edge vortex does not exhibit major modifications. Our results provide a better understanding of the flow in fluvial and coastal environments, informing engineering solutions aimed at containing the water flow and protecting banks and coasts from erosion.   
\end{abstract}

\backsection[Keywords]{Turbulent Flows, Geophysical and Geological Flows, Multiphase flow}

\input{intro}
\input{methods}
\input{setup}
\input{results}
\input{conclusions}

\backsection[Acknowledgements]{
The research was supported by the Okinawa Institute of Science and Technology Graduate University (OIST) with subsidy funding to M.E.R. from the Cabinet Office, Government of Japan. M.E.R.~also acknowledges funding from the Japan Society for the Promotion of Science (JSPS), grants 24K00810 and 24K17210. The authors acknowledge the computer time provided by the Scientific Computing and Data Analysis section of the Core Facilities at OIST and the computational resources offered by the HPCI System Research Project with grants hp220402, hp240006, hp250035. E.T.~acknowledges the support of the OIST Research Internship Program.}

\backsection[Declaration of interests]{The authors report no conflict of interest.}

\backsection[Data availability statement]{The plotted data are available at \url{XXX INSERT URL XXX}.}

\backsection[Author ORCIDs]{\\
Giulio Foggi Rota \orcidA{}: \url{https://orcid.org/0000-0002-4361-6521} \\
Elisa Tressoldi \orcidB{}: \url{https://orcid.org/0009-0003-6055-6836} \\
Francesco Avallone \orcidC{}: \url{https://orcid.org/0000-0002-6214-5200}\\
Marco Edoardo Rosti \orcidD{}: \url{https://orcid.org/0000-0002-9004-2292}
}

\appendix
\input{validation}

\bibliographystyle{jfm}

\end{document}

%% file: intro.tex
\section{Introduction} \label{sec:intro}

The presence of aquatic vegetation on fluvial and marine beds drastically affects the motion of waters above them \citep{nepf-2012-2}, with significant implications for turbulence research \citep{finnigan-2000} and coastal protection \citep{zhang-nepf-2021, mcwilliams-2023}. 

When multiple plants are arranged in an array, a vegetation canopy --- or simply \textit{canopy} --- is attained. The features of the turbulent flow established by a high--Reynolds number current impinging on the plants vary dramatically according to the tightness of their packing \citep{poggi-etal-2004,monti-etal-2022}, resembling those observed over a rough surface in a sparse scenario \citep{sharma-garciamayoral-2018} and those of a shear layer in dense conditions \citep{sharma-garciamayoral-2020-2}. In these latter conditions, the mixing properties of the current are enhanced \citep{poggi-katul-albertson-2004, ghisalberti-2010,mossa-etal-2017} with a positive effect on the transport and dispersal of suspended particles and nutrients. Contemporarily, the canopy shields the lower regions of the flow, depleting the lifting of sediments and limiting the erosion of the bed \citep{zhao-nepf-2021}. The suspension of sediments in the water gives rise to density imbalances responsible for the onset of gravity-driven currents \citep[i.e., turbidity currents][]{meiburg-kneller-2010}, which are recently being investigated experimentally in the context of canopy flows \citep{meredith-mcconnochie-nokes-2022, meredith-etal-2025}. 

Marine vegetation exhibits a great variety of morphologies: mangroves (\textit{Rhizophora mangle}) have stiff cylindrical roots emerging from the water, seagrass (\textit{Zostera marina}) is made of flexible blades, while kelp (\textit{Macrocystis pyrifera}) has a more complex shape with stiffer branches and looser leaves. It is thus not surprising that, depending on the species of plants involved, different canopy flows are attained \citep{ghisalberti-nepf-2006,nepf-2012-1}. 

While the first, seminal investigations \citep[e.g.][]{taddei-manes-ganapathisubramani-2016, monti-omidyeganeh-pinelli-2019} considered canopies made by rigid cylinders, the interest towards more complex configurations is growing \citep{fu-etal-2023}. Unique structural behaviours are observed when considering the motion of flexible stems. Already an isolated stem, depending on its rigidity, exhibits different regimes of motion when exposed to an incoming turbulent flow \citep{foggirota-etal-2024} or to a surface wave \citep{foggirota-chiarini-rosti-2025}. Yet, when organised in a canopy \citep{he-liu-shen-2022}, multiple flexible stems might additionally give rise to a collective waving motion known under the name of \textit{honami/monami} \citep{finnigan-1979}. Monami waves are now understood to be the fingerprint of the turbulent structures populating the shear--dominated region at the canopy tip \citep{monti-olivieri-rosti-2023}, and they are observed regardless of the regime of motion of the individual stems \citep{foggirota-etal-2024-2}. From an applicative perspective, the collective motion of ciliated meta--surfaces can be programmed electronically for micro--fluidic manipulation purposes \citep{wang-etal-2022-2}.

A further step towards the investigation of more realistic scenarios in experiments and simulations is achieved breaking the homogeneity of the canopy array along (one of) the wall parallel directions. Aquatic vegetation, in facts, is often zonal and organised in well-delimited patches, with abrupt transitions to non--vegetated regions of the bed. Novel vortical structures form at the canopy edge, impacting the {mean flow} and the local structure of turbulence. The \textit{edge-effects} clearly vary in the case of a streamwise \citep{nezu-onitsuka-2001,yan-etal-2016,yan-etal-2022-1,winiarska-etal-2023,winiarska-liberzon-vanhout-2024} or spanwise--oriented \citep{white-nepf-2007, moltchanov-etal-2015} vegetation discontinuity, with even more complex interactions arising in the case of a fully tridimensional patch \citep{rominger-nepf-2011, yan-etal-2022-2, park-nepf-2025}. 

The flow aligned to the canopy edge, in particular, draws much interest due to its analogy with the conditions attained at the river banks \citep{unigarovillota-etal-2023}, where the riparian vegetation slows the currents to a halt approaching dry land and attenuates incoming tidal bores \citep{serra-oldham-colomer-2018}. Large--scale vortices transfer momentum from the outer free waters to the canopy interior through turbulent diffusion, while momentum is consumed by small--scale circulations within the canopy \citep{yan-etal-2023}. Vortex-based \citep{white-nepf-2008} and drag-based \citep{jia-etal-2022,song-etal-2024} models have been employed to investigate the complex exchange of momentum between the vegetated and the non-vegetated regions of the so--called \textit{partially--obstructed channels} (i.e., open channel flows with the bed partly occupied by vegetation, with a sharp and elongated vegetation discontinuity along the streamwise direction), yet, to date, only experiments have been able to tackle the fluid motions at the stem level.

Few experiments and no simulations have integrated realistic features of the vegetation like the flexibility of the stems in the investigated setup. Recent advances \citep{caroppi-etal-2021} suggest that incorporating natural plant properties provides an improved description of the shear penetration and momentum exchange processes associated to the large-scale vortices. Thus, major developments in the simulation of flow processes in natural vegetated settings are possible only when truly acknowledging the morphological and bio--mechanical properties of the vegetation.

In this work we perform and analyse direct numerical simulations (DNSs) of the turbulent flow along the edge of a vegetation canopies made by individually resolved flexible stems, for two different values of their rigiditiy. Comparing the results to those attained in a non-vegetated channel and over a homogeneous canopy \citep{foggirota-etal-2024-2}, we infer the dominant mechanisms responsible for the exchange of momentum between the canopy and the open--flow region, further elucidating the effect of the stem rigidity on the resulting flow. Building on top of our former understanding \citep{foggirota-etal-2024}, we delve into the dynamical response of the flexible stems, highlighting its variation with the distance from the canopy edge. 

The remaining part of this paper is thus organised as follows: \S\ref{sec:methods} accurately describes the methods of our numerical experiments, performed in the setup reported in \S\ref{sec:setup}. The results are presented in \S\ref{sec:results}, where we start describing the {mean flow} \S\ref{sSec:meanFlow}, later characterising the turbulent fluctuations and the stress balances \S\ref{sSec:flucts}. We thus identify the most relevant turbulent structures and look at the events they induce at the top and at the edge of the canopy \S\ref{sSec:structs}. The discussion ends with a survey of the stem deformation and dynamics \S\ref{sSec:stems}. A critical overview of the results is found in \S\ref{sec:conclusions}, along with the concluding remarks. 

%% file: methods.tex
\section{Methods} \label{sec:methods}

\begin{figure}
\includegraphics[width=.95\textwidth]{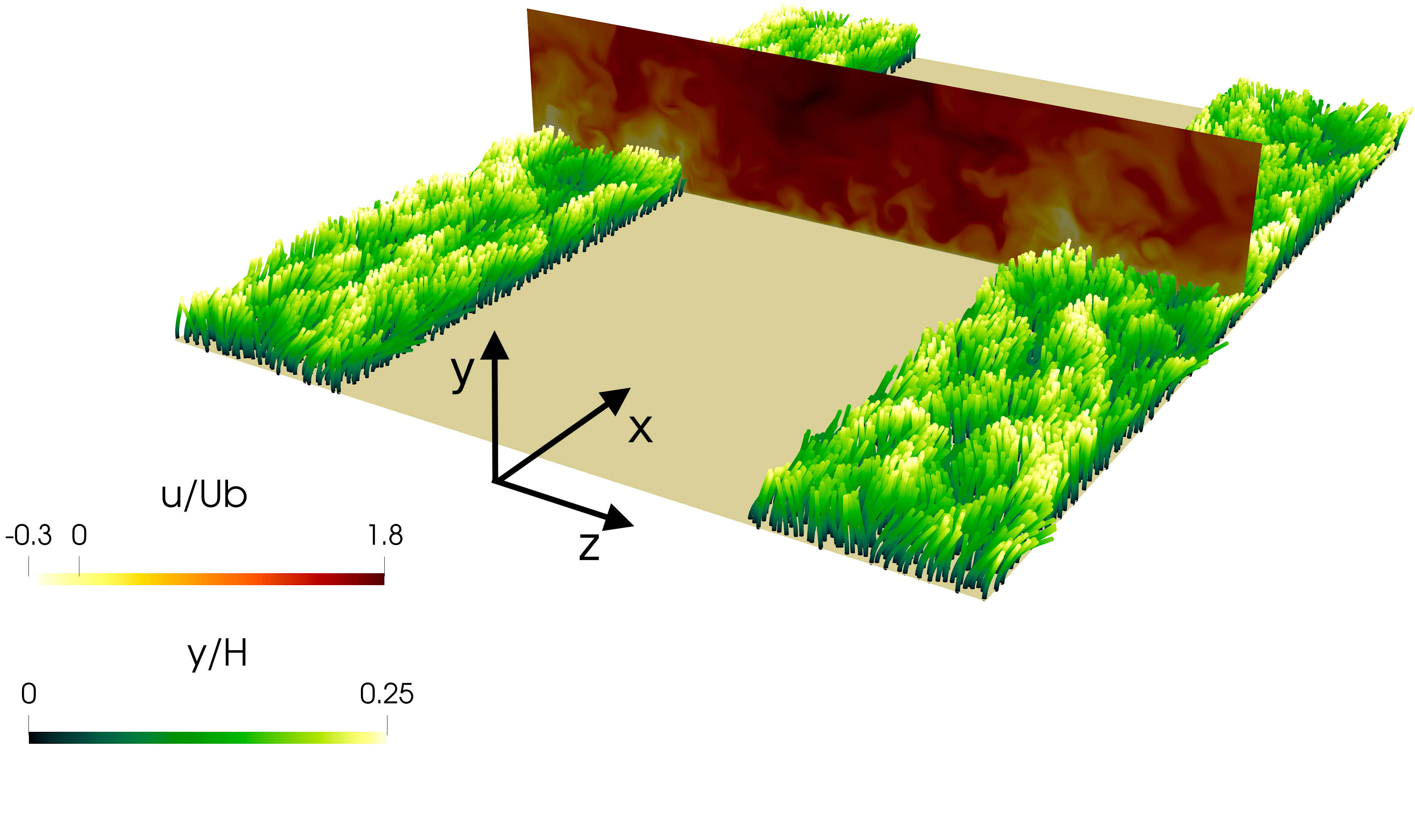}
\caption{Visualisation of the computational domain for the most compliant stems of our study ($Ca=100$). The flow is forced along the periodic $x$--axis, while the bottom wall and the top free-slip surface are orthogonal to the $y$--axis. The slice reports an instantaneous visualisation of the streamwise velocity fluctuations, $u'$. Flexible stems are vertically clamped on half of the bottom wall, leaving a non-vegetated gap of constant width in the periodic $z$--axis. Shades of green denote their elevation.}
\label{fig:setup} 
\end{figure}

Our DNSs are performed in a computational box with the $x$--axis aligned to the flow direction, the $y$--axis orthogonal to the plane where the canopy is anchored, and the $z$--axis oriented spanwise, in a right--handed convention. Flexible stems populate half of the bottom wall, equally parted between the left and the right sides as shown in figure~\ref{fig:setup}.

The flow of the incompressible Newtonian fluid is governed by the mass (\ref{eq:mass}) and momentum (\ref{eq:momentum}) balances. Let $\mathbf{u}(\mathbf{x},t)$ denote the velocity field and $p(\mathbf{x},t)$ the pressure field, each dependent on the spatial coordinates $\mathbf{x}$ and time $t$. Accordingly, the equations governing the dynamics of the fluid write
\begin{eqnarray}
   &\nabla \cdot \mathbf{u} = 0,
   \label{eq:mass}\\
   &\displaystyle \frac{\partial \mathbf{u}}{\partial t} + \nabla \cdot (\mathbf{u} \mathbf{u}) = - \frac{1}{\rho_f} \nabla p + \frac{\mu}{\rho_f} \nabla^2 \mathbf{u} + \mathbf{f}_s + \mathbf{f}_b,
   \label{eq:momentum}
\end{eqnarray}
where $\rho_f$ and $\mu$ are the volumetric density and dynamic viscosity of the fluid. No--slip and no--penetration boundary conditions are enforced at the bottom wall, while free-slip and no-penetration are enforced at the top surface of the domain. The $x$ and $z$ directions are treated as periodic. The force field $\mathbf{f}_s$, which accounts for the presence of the stems, is computed using a Lagrangian immersed boundary method (IBM) \citep{peskin-2002, huang-etal-2007, banaei-rosti-brandt-2020,olivieri-etal-2020-2}, as detailed below. In contrast, the force field $\mathbf{f}_b$ is applied uniformly over all grid points to ensure that the desired streamwise flow rate is attained at every time step, as introduced in the following section.

We discretise equations~(\ref{eq:mass},~\ref{eq:momentum}) on a staggered Cartesian grid using a second-order central finite difference scheme for both velocity and pressure. The grid consists of $N_x\times N_y\times N_z=1152\times 384\times 864$ points uniformly distributed along the periodic directions, while a non-homogeneous stretched distribution is employed along the $y$ axis to accurately capture the sharp velocity variation at the canopy tip. In particular, a finer and locally uniform resolution is used in the lower region of the domain --- comprising the canopy and the vegetation gap --- with a constant wall--normal spacing of $\Delta y/H = 0.002$ for $y/H \in [0.0,0.3]$, which then transitions smoothly to a spacing of $\Delta y/H = 0.004$ at $y/H =1$. Time integration is carried out using a second-order Adams-Bashforth scheme within a projection-correction framework \citep{kim-moin-1985}. The Poisson equation is solved efficiently via a Fast Fourier Transform (FFT)-based algorithm \citep{dorr-1970}, and the entire code is parallelised using the Message Passing Interface (MPI) together with the \textit{2decomp} library.

We model the stems as one-dimensional entities following a generalised version of the Euler-Bernoulli beam model, based on an extended version of the distributed-Lagrange-multiplier/fictitious-domain (DLM/FD) formulation introduced by \cite{yu-2005}. Denoting by $\mathbf{X}(s,t)$ the position of a point on the neutral axis of a stem, parametrised by the curvilinear abscissa $s$ and time $t$, the equations governing the dynamics of the structure write
\begin{eqnarray}
    &\Delta \Tilde{\rho} \displaystyle \frac{\partial^2 \mathbf{X}}{\partial t^2} = 
    \displaystyle \frac{\partial}{\partial s}\left(T\displaystyle \frac{\partial \mathbf{X}}{\partial s}\right) - \gamma \displaystyle \frac{\partial^4 \mathbf{X}}{\partial s^4} - {\mathbf{F_{IBM}}}, 
    \label{eq:eulerBernoulli}\\
    &\displaystyle \frac{\partial \mathbf{X}}{\partial s} \cdot \displaystyle \frac{\partial \mathbf{X}}{\partial s} = 1,
    \label{eq:inextensibility}
\end{eqnarray}
where $T$ is the tension enforcing inextensibility and $ {\mathbf{F_{IBM}}}$ is the force acting on the stems, computed by the Lagrangian IBM to couple them with the fluid, as described later. These equations (\ref{eq:eulerBernoulli},\ref{eq:inextensibility}) are supplemented with appropriate boundary conditions. Specifically, we impose $\mathbf{X}\rvert_{s=0}=\mathbf{X_0}$ and ${\partial\mathbf{X}}/{\partial s}\rvert_{s=0}=(0,1,0)$ at the clamp, and at the free end ($s=h$) we require ${\partial^3\mathbf{X}}/{\partial s^3}\rvert_{s=h}={\partial^2\mathbf{X}}/{\partial s^2}\rvert_{s=h}=\mathbf{0}$ along with $T\rvert_{s=h}=0$. The discretization of equations~(\ref{eq:eulerBernoulli},~\ref{eq:inextensibility}) follows the approach by \cite{huang-etal-2007}, expect for the bending term which is treated here implicitly as in \cite{banaei-rosti-brandt-2020} to allow for a larger time step. Each stem is discretised into a sequence of $32$ Lagrangian points, evenly spaced from the root to the tip.

In the absence of external forcing, a normal mode analysis of the structural equations yields the natural frequency 
\begin{equation}
f_{nat}=\left(\frac{\beta_1}{2\pi h^2}\right)\sqrt{\frac{\gamma}{\tilde{\rho}_s}}.
\end{equation}
Here, the stem density per unit length is defined as $\tilde{\rho}_s = \rho_f / (\pi r^2) + \Delta \tilde{\rho}$, and $\beta_1$ is a coefficient, approximately equal to $3.516$, determined from the analysis. The value of $f_{nat}$ is critical in determining the dynamical response of the stems to the fluid \citep{foggirota-etal-2024, foggirota-etal-2024-2}. 

Although the flexible stems, which sway in the flow, may collide with the wall or with one another, previous extensive testing \citep{monti-olivieri-rosti-2023,foggirota-etal-2024-2} has shown that stem-to-stem collisions exert only a weak influence on both the stems and the fluid dynamics, while stem-to-wall interactions occur only at $Ca$ values larger than those considered here. Both phenomena are thus neglected in the following.

The fluid--structure coupling is achieved by spreading the force distribution computed via the Lagrangian IBM over the Eulerian grid points, thereby enforcing the no-slip condition $\partial\mathbf{X}/\partial t=\mathbf{u}[\mathbf{X}(s,t),t]$ at the Lagrangian points representing the stems. The magnitude of the force ${\mathbf{F_{IBM}}}$ exerted by the fluid on the structure is taken to be proportional to the difference between the velocity of the structure and that of the fluid, interpolated at the structure points, $\mathbf{u}_{\mathbf{\rm{IBM}}}$. Accordingly, we express
\begin{equation}
{
\mathbf{F_{IBM}}= \beta \left(\mathbf{u}_{\mathbf{\rm{IBM}}} - \frac{\partial \mathbf{X}}{\partial t}\right),
}
\end{equation}
where $\beta$ is a tuned coefficient, set here equal to $10$. Finally, $\mathbf{F}_{\mathbf{\rm{IBM}}}$ is distributed to the nearby grid points to compute the reaction force on the fluid,
\begin{equation}
\mathbf{f}_s=\int_\Gamma \mathbf{F}_{\mathbf{\rm{IBM}}}(s,t)\delta(\mathbf{x}-\mathbf{X}(s,t))ds,
\end{equation}
with $\Gamma$ denoting the support of the IBM. Thus, the interface between the fluid and the filaments is not sharply defined but is instead spread over the IBM support via a window function that determines the effective diameter of the filaments.

We utilise our extensively validated solver \href{https://www.oist.jp/research/research-units/cffu/fujin}{Fujin} \citep{olivieri-etal-2020-2, brizzolara-etal-2021, olivieri-mazzino-rosti-2022, monti-olivieri-rosti-2023, foggirota-etal-2024, foggirota-etal-2024-2} to integrate in time the coupled equations that describe the motion of the fluid and of the stems. The very same methods and discretisation used here were adopted to investigate a full canopy in former works \citep{monti-olivieri-rosti-2023,foggirota-etal-2024-2}. There, the suitability of our numerical grid to correctly describe the dynamics of the flow and of the stems without introducing spurious effects is extensively assessed, along with the accuracy of the dynamical coupling between the fluid and the structure. In appendix \ref{app:validation}, we report for completeness the comparison between the {mean flow} profiles over a full canopy predicted by our code and those measured in experiments \citep{shimizu-etal-1992}, as previously published \citep{monti-olivieri-rosti-2023}.

%% file: setup.tex
\subsection{Setup} \label{sec:setup}

Our computational box has size $L_x \times L_y \times L_z = 2\pi H \times H \times1.5\pi H$, compatibly with the established guidelines for this kind of simulations \citep{sathe-giometto-2024}. 
Such domain is large enough to accommodate the streamwise-elongated turbulent structures generated at the canopy tip, while also fitting the spanwise-elongated rollers that populate the flow over the canopy \citep[see \S\ref{sSec:structs} and][]{bailey-stoll-2013, monti-olivieri-rosti-2023}.
The vertical dimension of the domain is instead likely to influence the results. In fact, our simulations focus on a submerged canopy rather than one exposed to a boundary layer --- which would necessitate a considerably larger domain. In simulations of submerged canopies, it is standard practice not to extend the fluid interface far above the canopy tip, but rather to approximate it as a free-slip surface \citep{he-liu-shen-2022, lohrer-frohlich-2023}; consequently, the distance from the bottom wall becomes a parameter. In this work, we follow this approach.

A constant flow rate is maintained dynamically adjusting the homogeneous body force $ \mathbf{f}_{b}$ imposed along the positive $x$--direction, so that the bulk Reynolds number has a constant value $\Rey_b=U_b H \rho_f/ \mu = 5000$. $U_b$ denotes the average streamwise velocity in the domain.

The 7776 flexible stems constituting the canopy have length $h = 0.25 H$ and radius $r \approx 0.01 H$. They are organised in a single patch, longitudinally split by the periodic boundary condition, so that they appear equally to the left and to the right of the domain, centred about a vegetation--gap of width $L_z/2$. This distance is sufficient to ensure fully--developed flow conditions at the middle of the gap when almost--rigid stems are considered (see \S\ref{sSec:meanFlow}). The vegetated region of the bed is regularly divided into squared tiles of edge $\Delta S / H \approx 0.044$, and a stem is clamped within each of them, orthogonal to the to the bottom wall, at a position randomly sampled from a uniform distribution to prevent preferential flow channeling effects.

{The solid volume fraction of the vegetation in the whole domain is $\Phi_v = \frac{N_s V_s}{L_x L_y L_z} \approx 2\%$, where $N_s$ is the total number of stems and $V_s = \pi r^2 h$ is their individual volume. $\Phi_v$ increases to $4\%$ if the vegetation gap is excluded from the estimate.  
This measure is closely related to the solidity, typically denoted $\lambda$, indicating the vegetation density \citep{monti-etal-2020,monti-etal-2022}.  
A homogeneous canopy made of vegetation with the same geometrical features considered here, such as that investigated by \cite{foggirota-etal-2024-2}, is characterised by $\lambda = 2 r h / \Delta S^2 \gg 0.1$. As such, it lies in the so-called dense regime, where vegetation has a profound effect on the flow. We thus choose the simulation parameters to ensure consistency with this regime, maximising the vegetation’s impact on the flow \citep{nepf-2012-1,monti-etal-2020,monti-etal-2022,foggirota-etal-2024-2}.} 

The dynamical properties of the stems are set by their bending rigidity $\gamma$ and their linear density difference from the surrounding fluid $\Delta \tilde{\rho}$. Throughout the investigation we set $\Delta \tilde{\rho}/(\pi r^2 \rho_f) = 0.27$, corresponding to slightly negatively buoyant stems, and explore two different values of $\gamma$ yielding a Cauchy number $Ca = \rho_f r h^3 U_b^2/\gamma \in \{10, 100\}$. $Ca$ represents the ratio between the deforming force exerted by the fluid and the elastic restoring force opposed by the stems, so that for $Ca =10$ the stems are rather stiff and straight, and the natural dynamics is expected to dominate their motion \citep{rosti-etal-2018}. For $Ca =100$, instead, the stems are more compliant to the incoming flow (see figure~\ref{fig:setup}). 

The parameter choice described above ensures exact correspondence with selected cases from our former investigations of flexible canopies extended indefinitely along the wall parallel directions \citep{foggirota-etal-2024-2}. {In particular, we ensure the matching of the solidity $\lambda$, adjusting the number of stems according to the bed coverage. The flow parameters are also matched with our simulations of an open channel with no vegetation, which serves as reference throughout this study together with the indefinite canopy ones (see table~\ref{tab:cases}).}

\begin{table}
    \centering
    \begin{tabular}{l c c c}
         & \textbf{Partially obstructed channel} & \textbf{Full canopy} & \textbf{Open channel} \\
        \midrule
        \textbf{Reference} & Present work  & \citet{foggirota-etal-2024-2}  & \citet{foggirota-etal-2024-2} \\
        {$\boldsymbol{Re_b}$} & $5000$  & $5000$  & $5000$ \\
        \textbf{Canopy coverage} & $50\%$  & $100\%$  & $0\%$ \\
        \textbf{$N_s$} & 7776 & 15552 & 0 \\
        {$\boldsymbol{Ca}$} & $10, 100$  & $10, 100$  & -- \\
        {$\boldsymbol{\Delta \tilde{\rho}/(\pi r^2 \rho_f)}$} & 0.27  & 0.27  & -- \\
        {$\boldsymbol{h/(2r)}$} & 25  & 25  & -- \\
        {$\boldsymbol{\Delta S/h}$} & 0.18  & 0.18  & -- \\
    \end{tabular}
    \caption{Simulations considered in our investigation}
    \label{tab:cases}
\end{table}

%% file: results.tex
\section{Results} \label{sec:results}

\subsection{{Mean flow}} \label{sSec:meanFlow}

\begin{figure}
\centering
\begin{subfigure}{\linewidth}
    \centering
    \includegraphics[width=.8\textwidth]{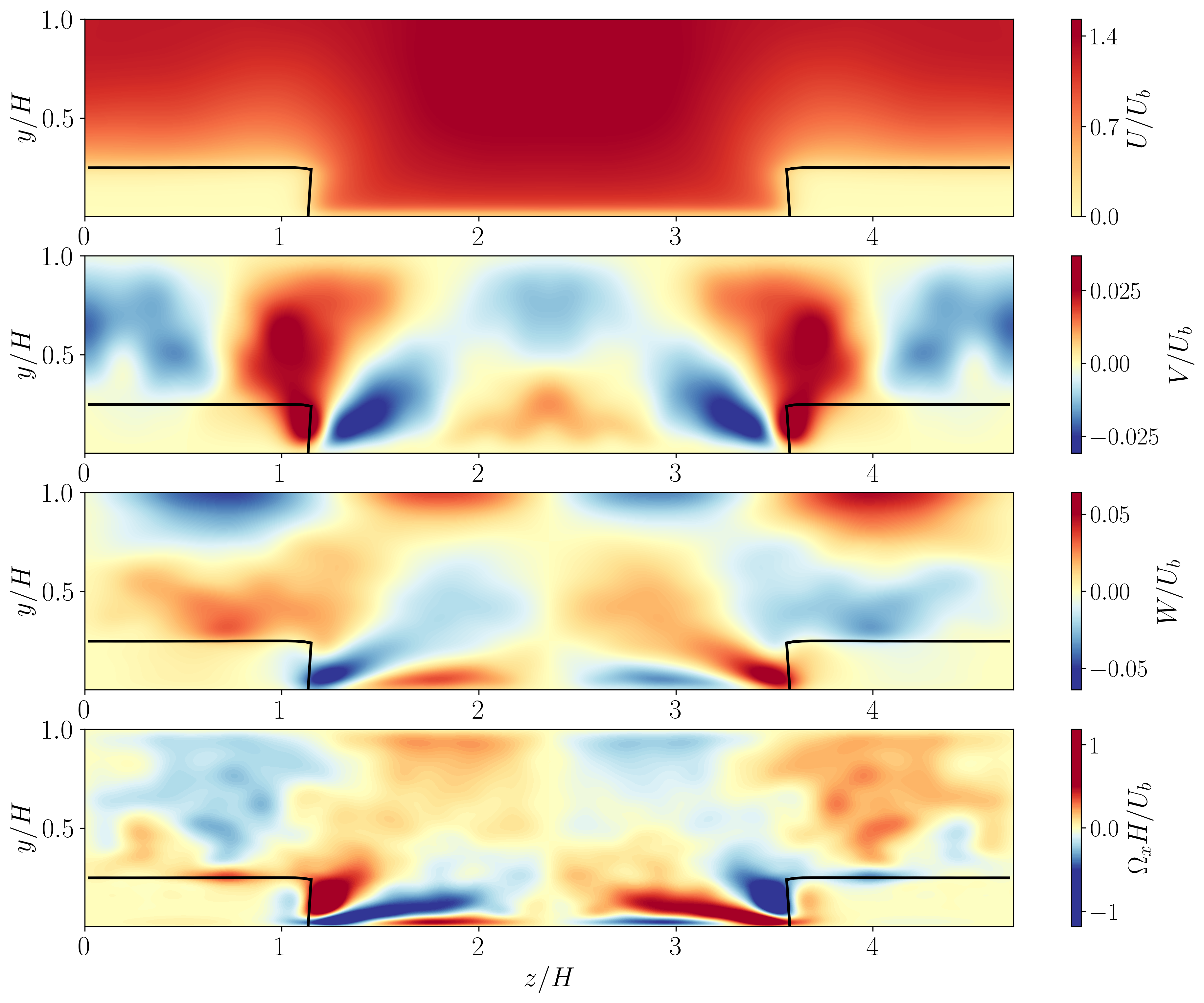}
    \put(-285,260){$Ca = 10$}
    \vspace{.5cm} 
\end{subfigure}
\begin{subfigure}{\linewidth}
    \centering
    \includegraphics[width=.8\textwidth]{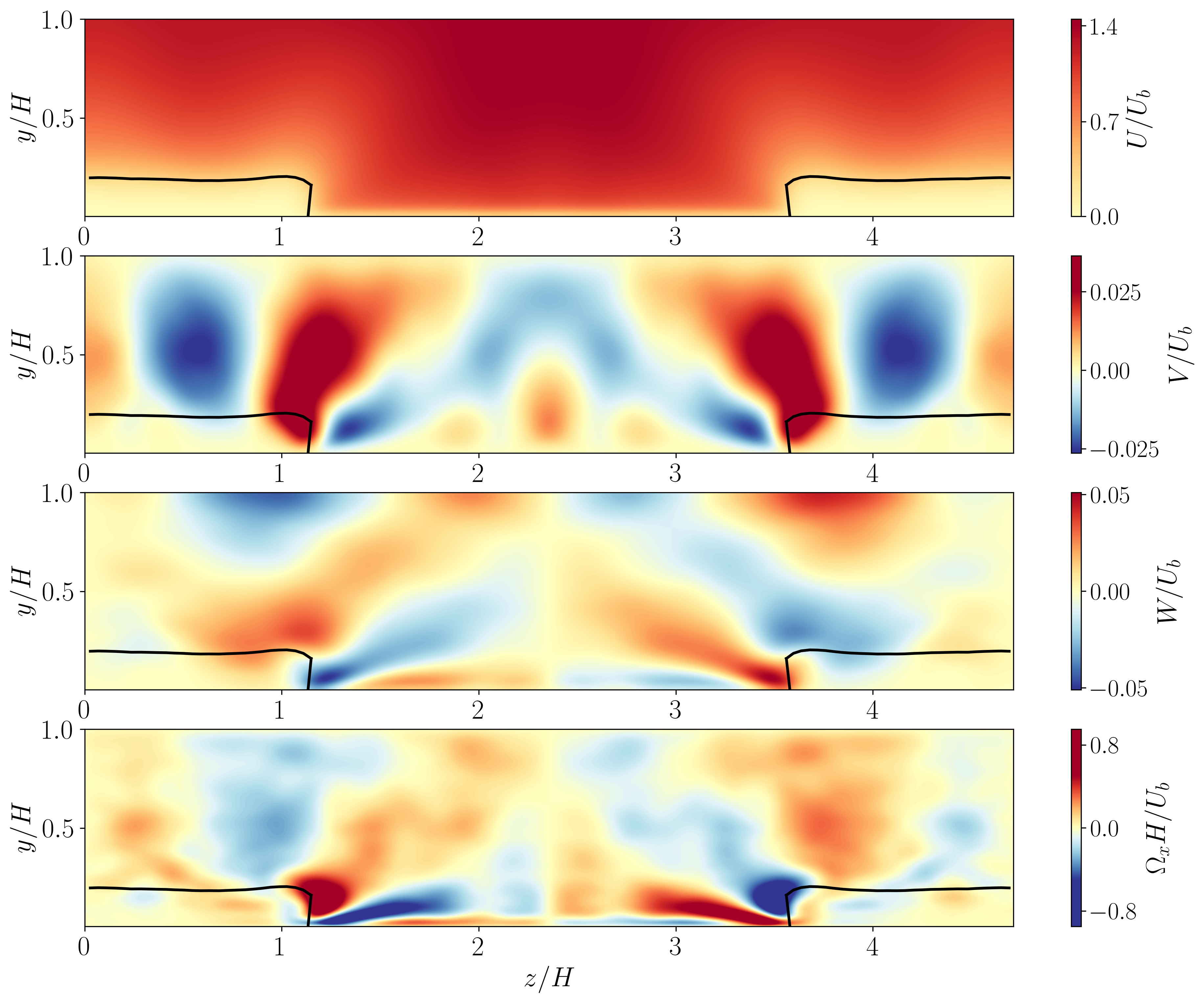}
    \put(-285,260){$Ca = 100$}
\end{subfigure}
\caption{Bidimensional {mean flow} established above and within the stems for the two different values of $Ca$ considered in our study. The total velocity $\mathbf{u}$ is averaged in time, along the $x$--axis, and made symmetric with respect to the middle of the vegetation gap, yielding $\mathbf{U}(y,z)=\{U,V,W\}$. The same averaging procedure is adopted to compute the {mean streamwise vorticity $\Omega_x$} and the envelope of the deflected stem tips, reported as a black line.}
\label{fig:meanFlow}
\end{figure}

{Given the turbulent nature of the flow fields, we decompose the velocity $\mathbf{u}$ into a mean component $\mathbf{U}$ and a fluctuating component $\mathbf{u'}$. 
Additionally, the {mean flow} along the canopy edge is known to develop a cellular structure \citep{unigarovillota-etal-2023} similar to what observed along streamwise-elongated roughness elements \citep{stroh-etal-2020}. Such structure emerges clearly from our simulations, and we thus better isolate it increasing our statistics and enforcing the symmetry exhibited by the system between the left and right halves of the domain.
This treatment is adopted only to enhance the intrinsic features of the {mean flow}, whereas the instantaneous fluctuations do not satisfy any symmetry constraint and their complex three-dimensional structure is fully retained.} The mean field $\mathbf{U}$ is thus attained averaging $\mathbf{u}$ in time, along the only homogeneous direction $x$, and exploiting symmetry with respect to the middle of the vegetation gap. Such peculiar average is denoted with angle brackets $\langle \cdot \rangle$ throughout the remaining part of the manuscript. At every time instant and at every point in space, 
\begin{equation}
	\mathbf{u} (\mathbf{x},t) = \langle \mathbf{u} \rangle(y,z) = \mathbf{U} (y,z) + \mathbf{u'} (\mathbf{x},t). 
\label{eq:decomp}
\end{equation}

We thus commence our analysis observing the fully bi-dimensional flow field $\mathbf{U}$ established above and between the stems {along with the mean envelope of the deflected stem tips, as reported in figure~\ref{fig:meanFlow}.
In fact, averaging in time and along the homogeneous $x$ direction the vertical coordinate of the tips belonging to stems located in tiles at the same spanwise location (i.e., within bins of size $\Delta S$ along the $z$-axis), we obtain a function $\langle \eta \rangle(z)$ representing the mean envelope of the deflected stem tips. 
We overlay this function to our plots (i.e., figures~\ref{fig:meanFlow}, \ref{fig:normalStresses} and \ref{fig:shearStresses}) to indicate, on average, the interface between the vegetation and the open flow above. 
At the lateral edge of the vegetated region, we close the envelope by drawing a straight line from the average position of the last stem tip down to its root.}

The streamwise component $U$ increases mowing away from the wall, reaching its maximum at the middle of the vegetation--gap; a slightly higher value is attained for the case at $Ca=10$ due to the higher blockage effect of the more rigid stems. 
{The dense canopy considered here significantly depletes the {mean flow} velocity within its interior, as typically observed in tightly packed porous media (see e.g.~\cite{rosti-cortelezzi-quadrio-2015}), yielding a sharp velocity increase emerging from its tip. The transition appears slightly less sharp in the more compliant case ($Ca=100$), where the outer flow penetrates further into the canopy array.}
The wall--normal component $V$ denotes the generation of an upwelling in correspondence of the interface between the vegetated and non--vegetated regions. A portion of the lifted fluid is deflected over the canopy, and thus impinges on the stems inducing their deflection (as clearly appreciated in the most flexible case, $Ca=100$). The remainder is deviated towards the centre of the vegetation--gap, from where it descends obliquely in the direction of the canopy. This is confirmed observing the spanwise component $W$, which together with $V$ highlights the formation of an intense circulating flow at the canopy edge. The streamwise oriented vortex formed at the canopy edge draws fluid inside the canopy from the side, and thus ejects it upward in the previously discussed upwelling. In the most flexible case, at $Ca=100$, the flapping motion of the stems is more intense and fluid is thus pumped from the canopy upward, enhancing the upwelling. A closer inspection of figure~\ref{fig:meanFlow} also reveals a region of weaker circulation moving away from the canopy towards the centre of the gap, counterrotating with respect to the canopy--edge vortex.
{This is further confirmed by the plots of the streamwise vorticity of the {mean flow} $\Omega_x$, which also corroborate the previous description of the canopy--edge vortex. Note that positive values of $\Omega_x$ correspond to clockwise circulation in the $y$–$z$ plane.}

Our observations are in qualitative agreement with the {mean flow} features reported by experiments conducted in partially--obstructed channels with a canopy of width comparable to {the one} of the gap \citep{yan-etal-2023}. Furthermore, the formation of an upwelling similar to that reported here is also observed at the interface between a smooth and a rough bed \citep[see case $h=0$ from][]{stroh-etal-2020}. Yet, it is interesting to notice that this scenario can be completely upset in the case of a less wide canopy, where the effect of the lateral confinement is felt more strongly at the interface between the vegetated and non--vegetated regions. In this case, experiments \citep{unigarovillota-etal-2023} report the formation of a downwelling at the interface, with fluid being drawn upward from deeper regions of the canopy. The direction of rotation of the canopy--edge vortex is therefore inverted with respect to what we report.  

\begin{figure}
    \includegraphics[width=.95\textwidth]{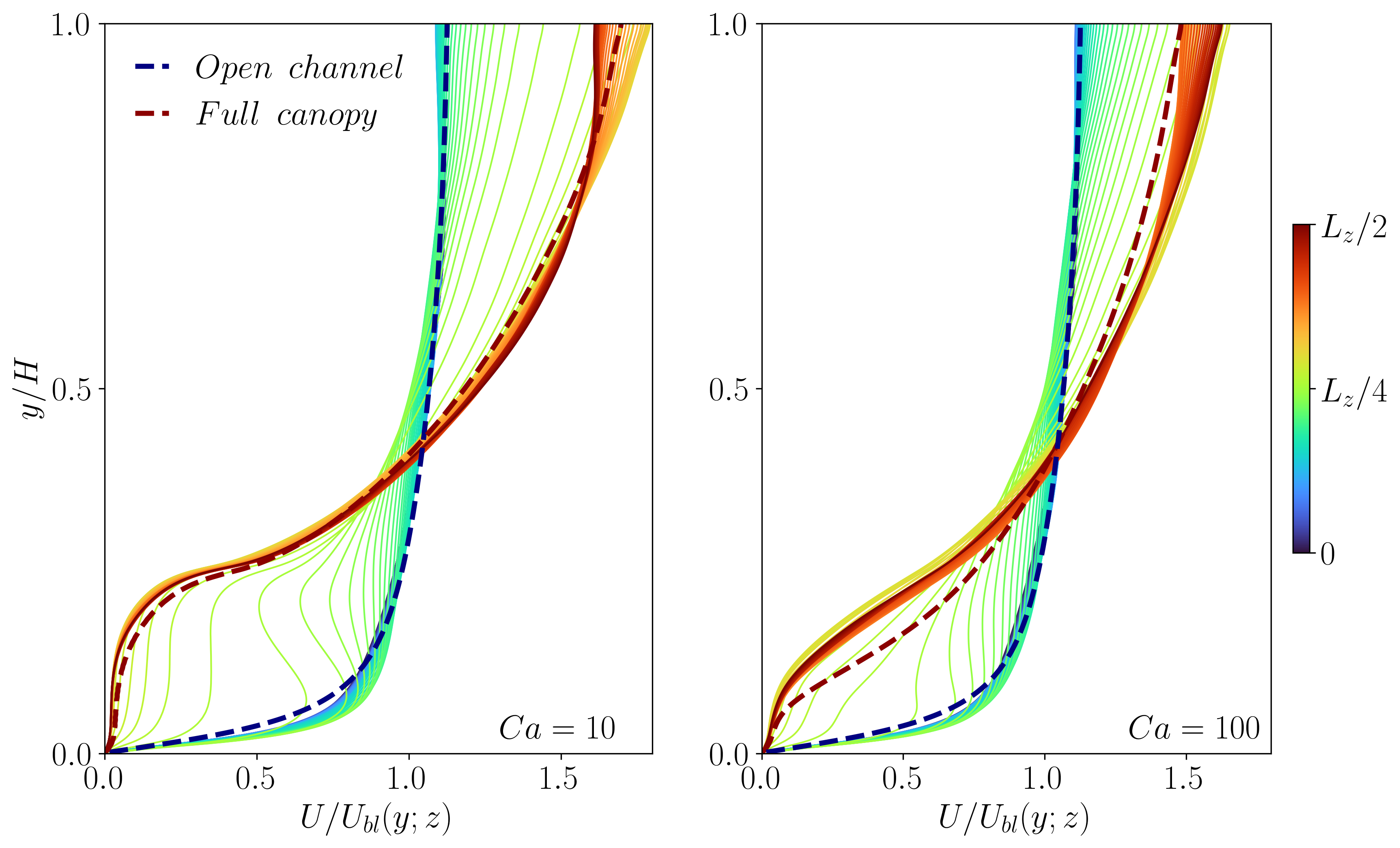}
    \caption{Wall-normal profiles of the mean streamwise velocity, normalised with the local bulk velocity $U_{bl}$ as defined in the text. Colours ranging from blue to red denote different positions along the $z$--axis. For comparison, we also report the profiles of an open channel (dashed blue line) and of full canopies (dashed red line) with the same parameters of those considered here, as measured in our former study \citep{foggirota-etal-2024-2}.}
    \label{fig:strProfiles}
\end{figure}

Next, we further characterise the streamwise component of the mean velocity field observing its wall--normal profiles at different positions along the $z$--axis, $U(y;z)$ as reported in figure~\ref{fig:strProfiles}. The profiles are normalised with respect to the local bulk velocity 
\begin{equation}
U_{bl}(z) = \int_0^H U(y,z) dy,
\end{equation}
and compared to those we measured in an open channel and in full canopies with the same parameters of those considered here \citep{foggirota-etal-2024-2}. Starting from the most rigid case, at $Ca=10$, we observe that the profile at the middle of the canopy resembles that found in the case of a full canopy. Analogously, the profile at the middle of the gap has a shape similar to what found in an open channel. We thus deduce that, in this case, the vegetated and non--vegetated regions are wide enough to attain fully--developed conditions at their respective middles, as far as the {mean flow} is concerned. At the interface between the two regions, located at $L_z/4$, the profile is non--monotonous, with the lower fluid moving faster than the one on top. This effect is a consequence of the bi-dimensionality of the {mean flow}, and follows from the presence of the canopy--edge vortex. High--speed fluid is drawn from the gap towards the lower region of the canopy by the vortex, through the vegetation interface. As it ascends in the upwelling, the fluid looses streamwise momentum due to the interaction with the stems and the non-monotonous trend of the profile is attained. Finally, upon exiting the canopy from the top, the fluid accelerates again due to the reduced drag.

The situation in the most flexible case, at $Ca=100$, appears qualitatively different. The profile at the middle of the gap still resembles that found in an open channel, but the one at the middle of the canopy deviates from that of a full canopy. Increasing the compliance of the stems, perturbations from the edge penetrate deeper in the canopy along the transversal direction and the {mean flow} is consequently depleted within the canopy region, compared to the full canopy case. The velocity far above the canopy is higher than in the full canopy merely due to mass conservation arguments. We also notice that, in this case, the monotonous growth of the profile is preserved at all spanwise locations. The increased updraft induced by the enhanced flapping motion of the stems, pumping the fluid upward, ensures that the high-speed fluid drawn by the canopy--edge vortex from the gap into the canopy emerges quicker, with a diminished loss of streamwise--momentum against the stems. Noticeably, looking back at figure~\ref{fig:meanFlow}, the canopy--edge vortex itself is not significantly affected by the increased flexibility of the stems. Yet, in this case the stems are deflected forward and sway in the flow (as discussed later \S\ref{sSec:stems}), exerting less drag on the fluid.

\subsection{Fluctuations and stress balances} \label{sSec:flucts}

\begin{figure}
\centering
\begin{subfigure}{\linewidth}
    \centering
    \includegraphics[width=.8\textwidth]{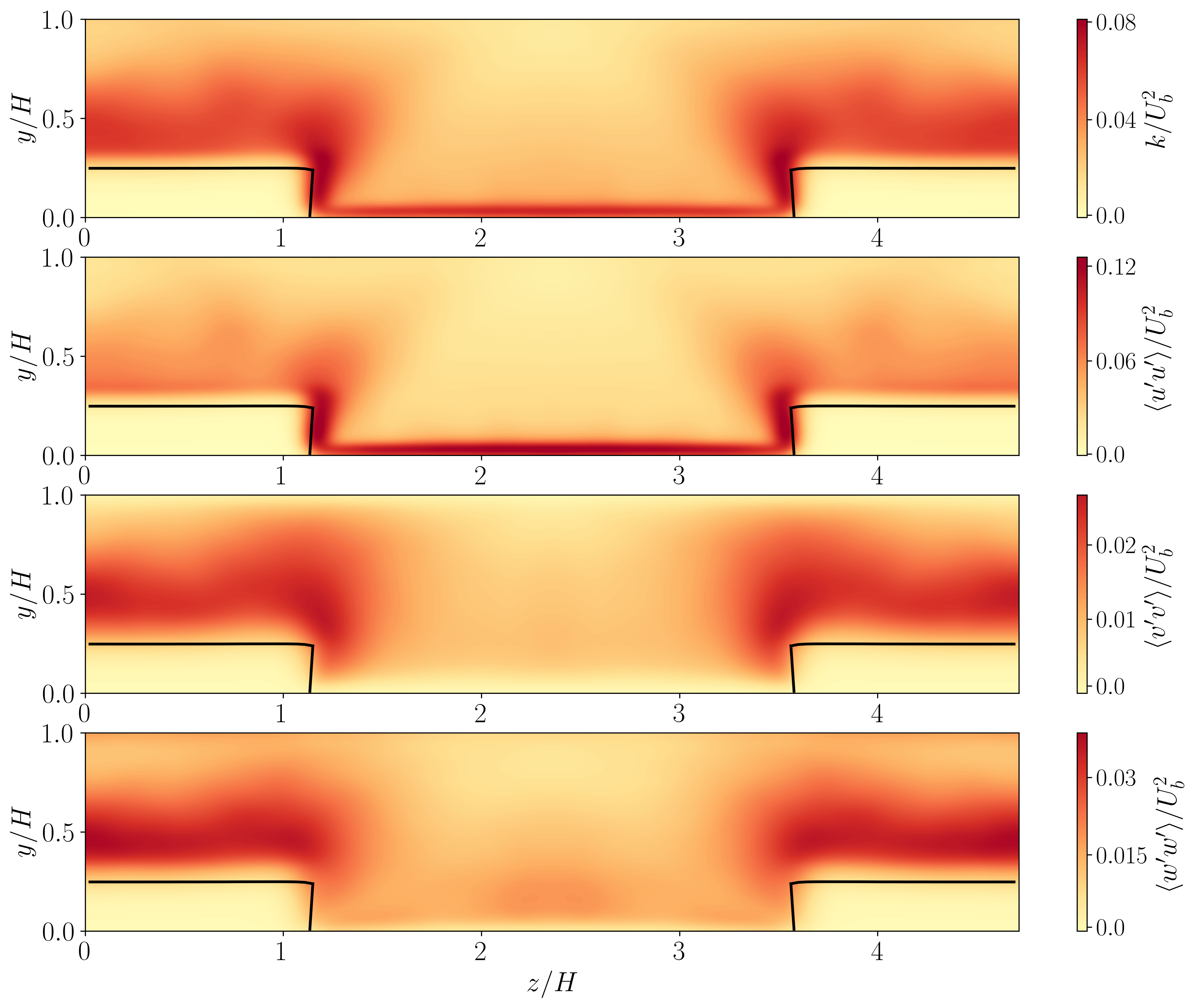}
    \put(-285,260){$Ca = 10$}
    \vspace{.2cm} 
\end{subfigure}
\begin{subfigure}{\linewidth}
    \centering
    \includegraphics[width=.8\textwidth]{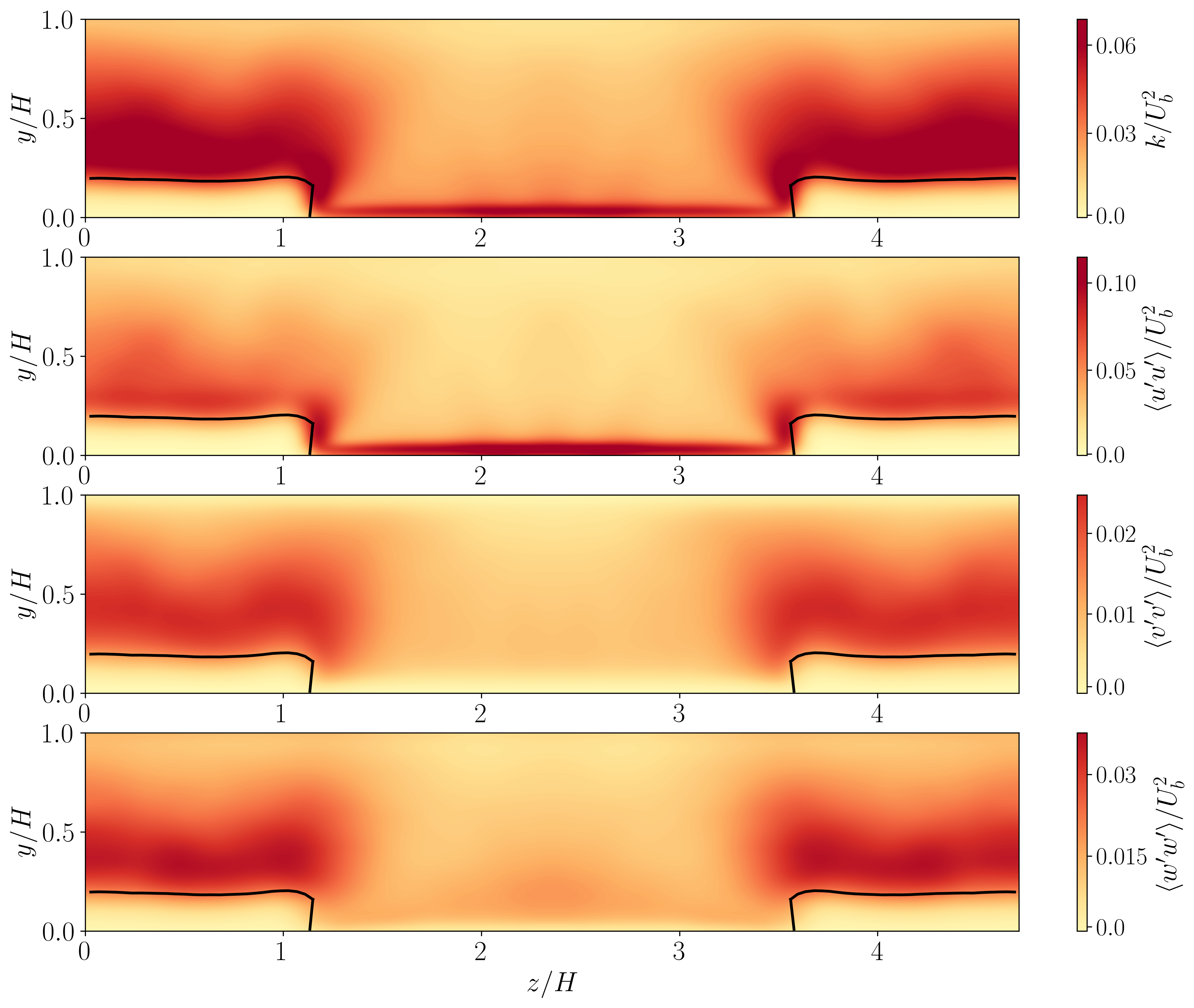}
    \put(-285,260){$Ca = 100$}
\end{subfigure}
\caption{Turbulent kinetic energy $k$ and normal components of the Reynolds' stress tensor in the spanwise $y$--$z$ plane, for the two different values of $Ca$ considered in our study. The fluctuations are averaged in time, along the $x$--axis, and made symmetric with respect to the middle of the vegetation gap. The same averaging procedure is adopted to compute the mean envelope of the deflected stem tips, reported as a black line.}
\label{fig:normalStresses}
\end{figure}

We now turn our attention to the fluctuating velocity field $\mathbf{u'} (\mathbf{x},t)$, introduced in equation~(\ref{eq:decomp}) {as the deviation from the symmetric mean flow specified in \S\ref{sSec:meanFlow}. Angled brackets, introduced there, denote the average employed to isolate it.} First, we focus on the spatial distribution of the turbulent kinetic energy $k = 0.5 (\langle u' u' \rangle + \langle v' v' \rangle + \langle w' w' \rangle)$ in the spanwise $y$--$z$ plane, as shown in the top panels of figure~\ref{fig:normalStresses}. In the most rigid case, $Ca=10$, $k$ appears more intense in proximity of the bottom wall within the gap, and along the interface between the vegetated and non-vegetated regions, attaining slightly lower values over the canopy. In the most flexible case, $Ca=100$, instead, the magnitude of $k$ over the canopy is comparable to that attained over the wall within the gap, suggesting that turbulence over the canopy is enhanced by the flapping motion of the stems.

Next, we observe separately the normal components of the Reynolds stress tensor $\langle \mathbf{u'}  \mathbf{u'} \rangle $ (in figure~\ref{fig:normalStresses}). Predictably, the most significant contribution to $k$ at the bottom wall within the gap comes from the streamwise component $\langle u' u' \rangle$, capturing the velocity fluctuations associated to the formation of low and high--speed streaks. Those are the fingerprint of the near--wall cycle  \citep{jimenez-pinelli-1999}, active there. $\langle u' u' \rangle$ appears depleted over the canopy, but still remains significant, compatibly with the formation of the large streak--like structures held responsible for the coherent swaying of the stems and the propagation of monami waves \citep{monti-olivieri-rosti-2023}. The dominant components over the canopy are nevertheless $\langle v' v' \rangle$ and $\langle w' w' \rangle$, symptomatic of the Kelvin--Helmholtz rollers populating the shear--layer \citep{finnigan-2000}. The magnitude of $\langle v' v' \rangle$ and $\langle w' w' \rangle$ drops moving towards the middle of the gap, with a slightly slower decay in the most rigid case, $Ca=10$, where the more rigid stems induce the formation of a sharper shear--layer. Thus, in this case, the rollers thus extend further from the canopy towards the gap.

\begin{figure}
\centering
\begin{subfigure}{\linewidth}
    \centering
    \includegraphics[width=.95\textwidth]{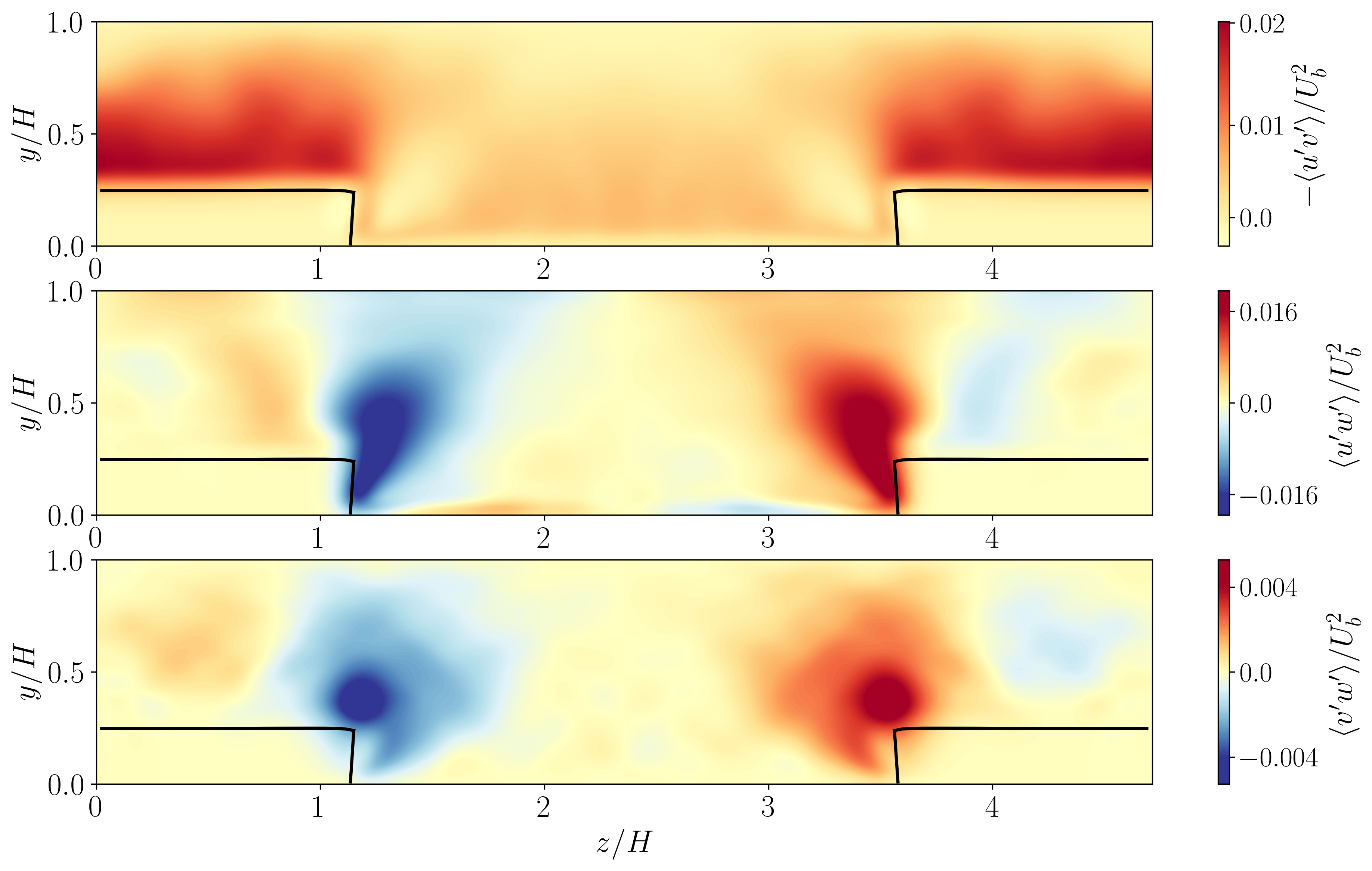}
    \put(-340,235){$Ca = 10$}
    \vspace{1.5cm} 
\end{subfigure}
\begin{subfigure}{\linewidth}
    \centering
    \includegraphics[width=.95\textwidth]{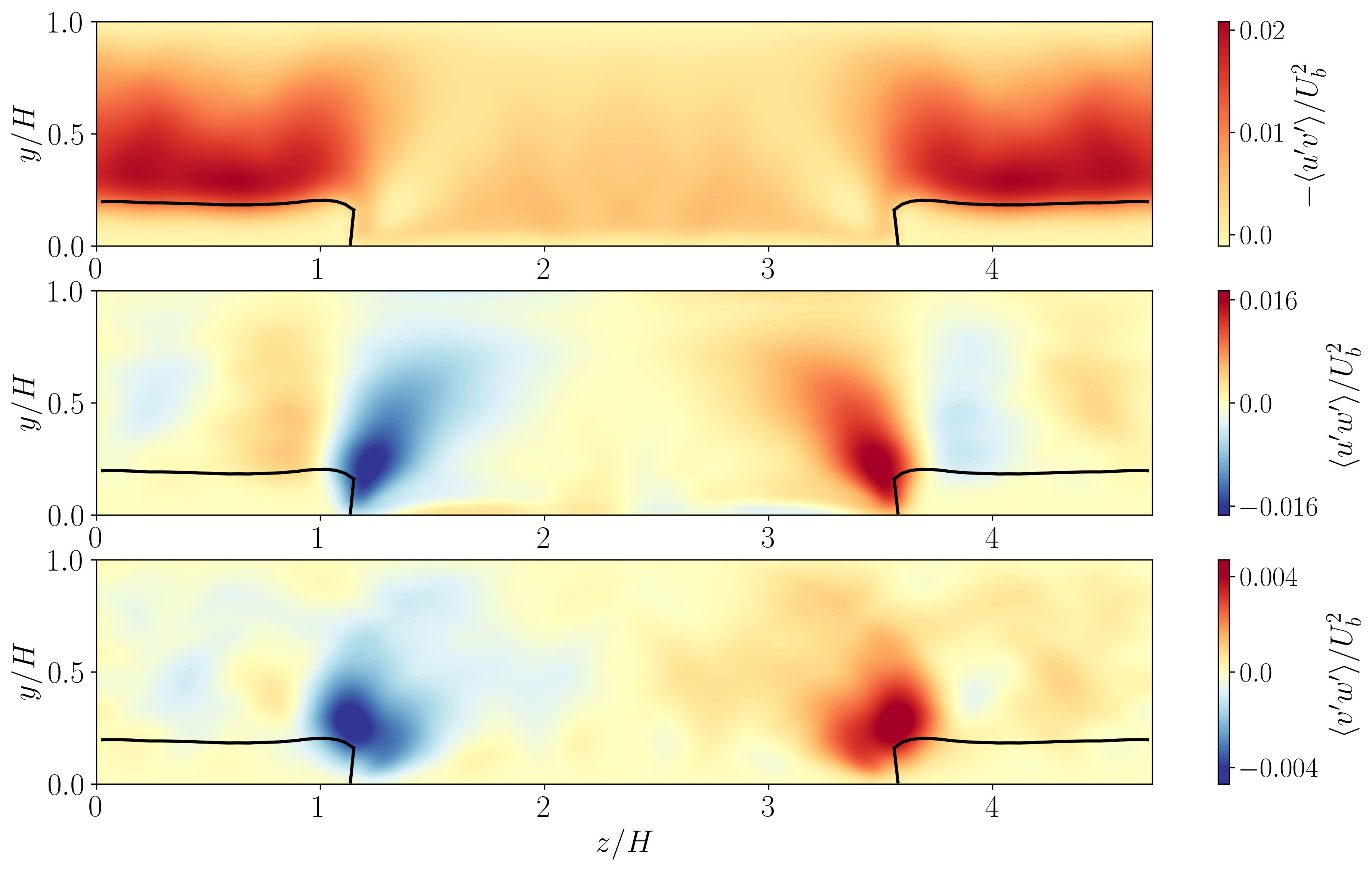}
    \put(-340,235){$Ca = 100$}
\end{subfigure}
\caption{Extra--diagonal components of the Reynolds' stress tensor in the spanwise $y$--$z$ plane, for the two different values of $Ca$ considered in our study. The fluctuations are averaged in time, along the $x$--axis, and made (anti--)symmetric with respect to the middle of the vegetation gap. The same averaging procedure is adopted to compute the mean envelope of the deflected stem tips, reported as a black line.}
\label{fig:shearStresses}
\end{figure}

We now look at the extra--diagonal terms of the Reynolds' stress tensor, as shown in figure~\ref{fig:shearStresses}. The $\langle u'v' \rangle$ component peaks above the canopy, where the Kelvin--Helmholtz rollers induce correlated fluctuations of the streamwise and wall--normal velocities. The $\langle u'w' \rangle$ component, instead, denotes regions where fluid with high and low streamwise momentum is exchanged between the canopy and the gap. High--speed fluid from the gap, entrained by the canopy--edge vortex, approaches the canopy from the side. Meanwhile, low--speed fluid emerging from the canopy deviates towards the gap above the canopy edge. These two mechanisms give rise to $u'w'$ events of the same sign, generating the regions of intense $\langle u'w' \rangle$ activity visible in the plots. In the bottom panels of figure~\ref{fig:shearStresses}, the joint fluctuations of $\langle v'w' \rangle$ peak close to the core of the canopy--edge vortex, revealing its unsteady nature. 

The role of the turbulent shear stresses is better understood within the context of the shear stress balance, which we therefore elucidate. 
Let us consider the momentum equation along the streamwise direction, 
\begin{equation}
    \frac{\partial u}{\partial t} + u \frac{\partial u}{\partial x} + v \frac{\partial u}{\partial y} + w \frac{\partial u}{\partial z} = - \frac{1}{\rho_f}\frac{\partial p}{\partial x} + \frac{\mu}{\rho_f} \left( \frac{\partial^2 u}{\partial x^2} + \frac{\partial^2 u}{\partial y^2} + \frac{\partial^2 u}{\partial z^2}\right) + f_{s,x} + f_{b,x}.
    \label{eq:momEq}
\end{equation}
For simplicity, we rename the force $- f_{s,x}$ exerted by the stems on the fluid $d$, for drag, and incorporate the forcing $f_{b,x}$ into a modified pressure gradient $\partial \tilde{p} / \partial x$. Introducing the decomposition in equation~(\ref{eq:decomp}), averaging, and reordering the terms, equation~(\ref{eq:momEq}) rewrites
\begin{multline}
   \frac{\mu}{\rho_f} \left( \frac{\partial^2 \langle u \rangle}{\partial y^2} + \frac{\partial^2 \langle u \rangle}{\partial z^2}\right) =  \frac{1}{\rho_f} \left \langle \frac{\partial \tilde{p}}{\partial x} \right \rangle + \langle u \rangle \frac{\partial \langle u \rangle}{\partial x} + \left\langle u' \frac{\partial u'}{\partial x} \right\rangle + \langle v \rangle \frac{\partial \langle u \rangle}{\partial y} + \\
    + \left \langle v' \frac{\partial u'}{\partial y} \right \rangle + \langle w \rangle \frac{\partial \langle u \rangle}{\partial z} + \left \langle w' \frac{\partial u'}{\partial z} \right \rangle + \langle d \rangle.
    \label{eq:momEq_avg}
\end{multline}
Integrating equation~(\ref{eq:momEq_avg}) between the bottom wall ($y = 0$) and a generic position $y$ in the wall--normal direction, there follows
 \begin{multline}
    \frac{\tau_w}{\rho_f} = \frac{\mu}{\rho_f} \frac{\partial \langle u \rangle}{\partial y} \bigg|_y + \frac{\mu}{\rho_f} \int_0^y \frac{\partial^2 \langle u \rangle}{\partial z^2} dy - \frac{1}{\rho_f} \left \langle \frac{\partial \tilde{p}}{\partial x} \right \rangle y - \int_0^y \left \langle u' \frac{\partial u'}{\partial x} \right \rangle dy -\\
    -\int_0^y \langle v \rangle \frac{\partial \langle u \rangle}{\partial y} dy - \int_0^y \left \langle v' \frac{\partial u'}{\partial y} \right \rangle dy - \int_0^y \langle w \rangle \frac{\partial \langle u \rangle}{\partial z} dy - \int_0^y \left \langle w' \frac{\partial u'}{\partial z} \right \rangle dy - \int_0^y \langle d \rangle dy,
    \label{eq:momEq_intY}
\end{multline}
where the shear stress at the wall, $\tau_w$, stems from 
\begin{equation}
   \frac{\mu}{\rho_f} \int_0^y \frac{\partial^2 \langle u \rangle}{\partial y^2} dy = \frac{\mu}{\rho_f} \frac{\partial \langle u \rangle}{\partial y} \bigg|_y - \frac{\mu}{\rho_f} \frac{\partial \langle u \rangle}{\partial y} \bigg|_0 = \frac{\mu}{\rho_f} \frac{\partial \langle u \rangle}{\partial y} \bigg|_y - \frac{\tau_w}{\rho_f}.
\end{equation}
Integrating again equation~(\ref{eq:momEq_avg}) along the wall--normal direction, this time between the bottom wall ($y = 0$) and the top surface ($y = H$), where the free--slip condition is enforced, we find
\begin{multline}
    \frac{\tau_w}{\rho_f} = \frac{\mu}{\rho_f} \int_0^H \frac{\partial^2 \langle u \rangle}{\partial z^2} dy - \frac{1}{\rho_f} \left \langle \frac{\partial \tilde{p}}{\partial x} \right \rangle H - \int_0^H \left \langle u' \frac{\partial u'}{\partial x} \right \rangle dy-\\
    - \int_0^H \langle v \rangle \frac{\partial \langle u \rangle}{\partial y} dy -\int_0^H \left \langle v' \frac{\partial u'}{\partial y} \right \rangle dy - \int_0^H \langle w \rangle \frac{\partial \langle u \rangle}{\partial z} dy - \int_0^H \left \langle w' \frac{\partial u'}{\partial z} \right \rangle dy - \int_0^H \langle d \rangle dy.
    \label{eq:momEq_intW}
\end{multline}
Equating the right--hand--sides of equations~(\ref{eq:momEq_intY}, \ref{eq:momEq_intW}), de facto removing $\tau_w$, we find the desired balance
\begin{multline}
   - \underbrace{\left \langle \frac{\partial \tilde{p}}{\partial x} \right \rangle (H - y)}_{\tau} =
   \underbrace{\rho_f \int_y^H \langle D \rangle dy}_{D} +
   \underbrace{\mu {\frac{\partial \langle u \rangle}{\partial y}}\bigg|_y}_{\tau_1} - 
   \underbrace{\mu \int_y^H \frac{\partial^2 \langle u \rangle}{\partial z^2} dy}_{\tau_2} + \\
   + \underbrace{\rho_f \int_y^H \left[ \langle v \rangle \frac{\partial \langle u \rangle}{\partial y} 
   + \langle w \rangle \frac{\partial \langle u \rangle}{\partial z} \right] dy}_{\tau_3} + 
   \underbrace{\rho_f \int_y^H \left[ \left \langle u' \frac{\partial u'}{\partial x} \right \rangle 
   + \left \langle v' \frac{\partial u'}{\partial y} \right \rangle 
   + \left \langle w' \frac{\partial u'}{\partial z} \right \rangle \right] dy}_{\tau_4},
   \label{eq:shearBalance}
\end{multline}
where the combination of all the different contributions balances the linear profile of the total shear stress $\tau_{tot}(y) =  - \left \langle {\partial \tilde{p}} /{\partial x} \right \rangle (H - y)$, with a linear slope set by the driving pressure--gradient. In particular, 
\begin{itemize}
	\item $D$ is the mean canopy drag;
	\item $\tau_1$ is the viscous shear stress;
	\item $\tau_2$ is the viscous diffusion of streamwise momentum in the spanwise direction;
	\item $\tau_3$ is the advection of streamwise momentum by the {mean flow} in the $y$--$z$ plane;
	\item $\tau_4$ is the turbulent shear stress. \\
\end{itemize}

\begin{figure}
\centering
\begin{subfigure}{\linewidth}
    \centering
    \includegraphics[width=\textwidth]{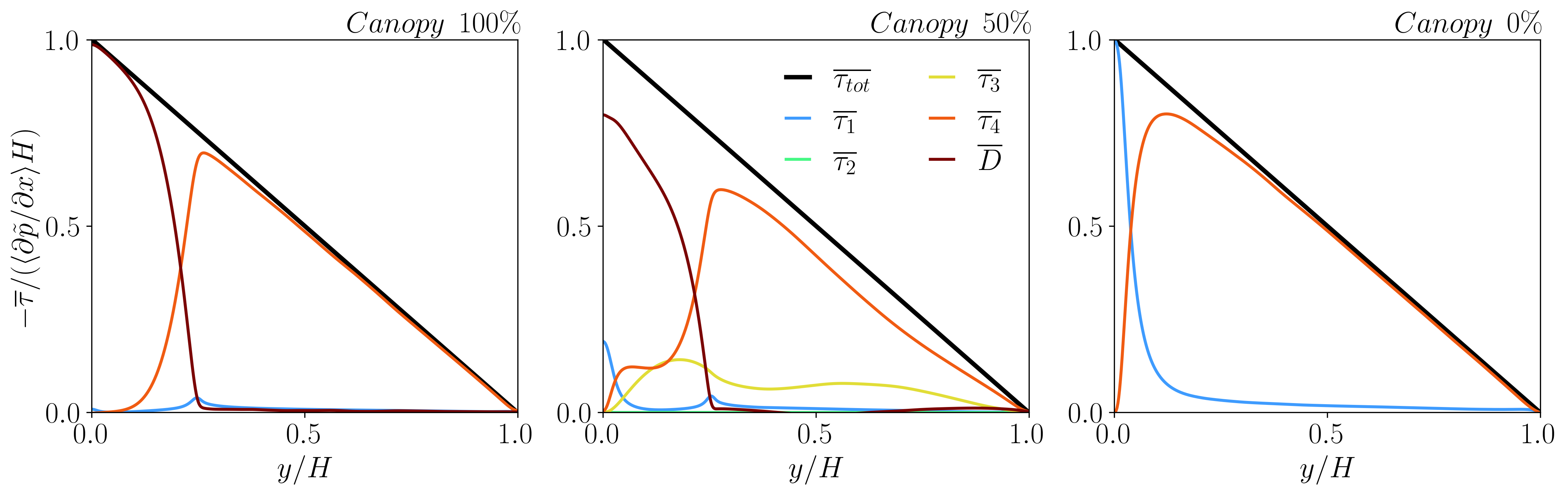}
    \put(-360,130){$Ca = 10$}
    \vspace{.25cm} 
\end{subfigure}
\begin{subfigure}{\linewidth}
    \centering
    \includegraphics[width=\textwidth]{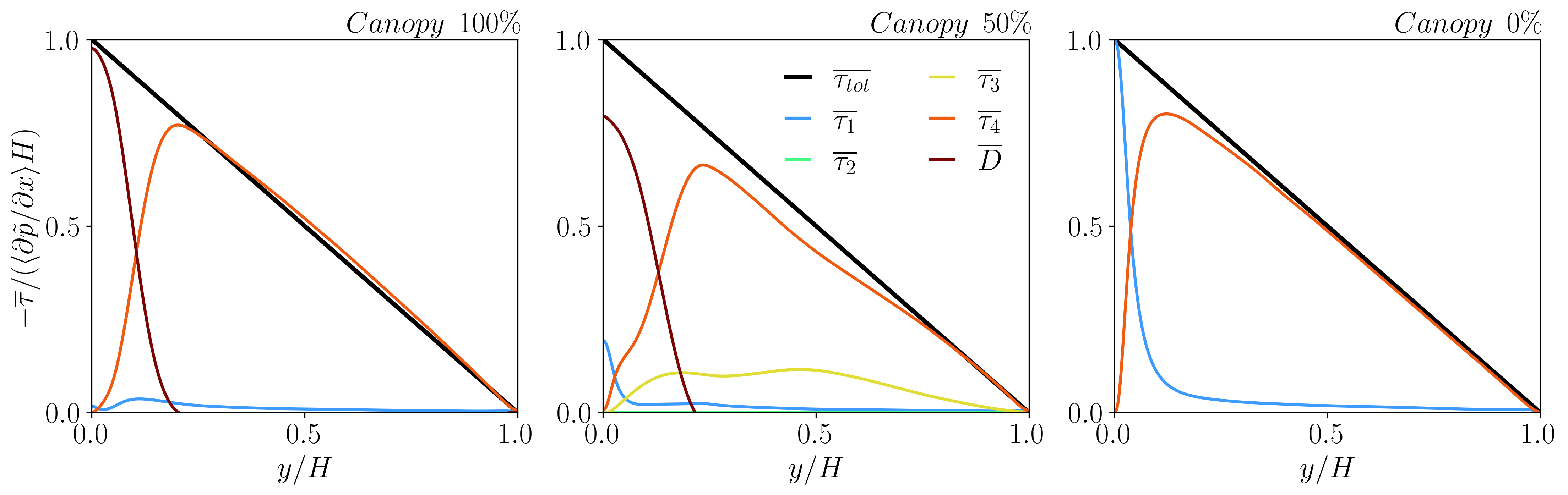}
    \put(-360,130){$Ca = 100$}
\end{subfigure}
\caption{Wall-normal profiles of the shear--balance terms computed from our simulations of a full canopy (canopy 100\%), a partially-obstructed channel (canopy 50\%), and an open channel (canopy 0\%), with the first and the last from our former investigation \citep{foggirota-etal-2024-2}. All flow parameters are matched between each set of simulations, and the different contributions are averaged along the periodic directions. We also ensure the matching of the structural parameters in the vegetated cases, for which we report two different values of stem flexibility. The combination of all the different contributions balances the linear profile of the total shear stress $\overline{\tau_{tot}}$, in particular: $\overline{\tau_1}$ is the viscous shear stress, $\overline{\tau_2}$ is the viscous diffusion of streamwise momentum in the spanwise direction, $\overline{\tau_3}$ is the advection of streamwise momentum by the {mean flow} in the $y$--$z$ plane, $\overline{\tau_4}$ is the turbulent shear stress, and $\overline{D}$ is the mean canopy drag.}
\label{fig:shearBalance}
\end{figure}

The contributions reported above are all functions of the spanwise and wall--normal coordinates. 
To ease the comparison with the homogeneous cases and with results typically reported in literature \citep{kim-moin-moser-1987}, we average them along the spanwise direction enforcing the periodicity of the domain. An over--bar denotes the newly introduced average operator.
Each average contribution is reported in figure~\ref{fig:shearBalance}, both at $Ca=10$ and $Ca=100$, and compared with the case of a full canopy and of an open channel with no vegetation. Note that, trivially, there is no average momentum diffusion nor advection ($\overline{\tau_2}=\overline{\tau_3}=0$) in the full canopy case, while in the case of an open channel also the canopy drag is null ($\overline{\tau_2}=\overline{\tau_3}=\overline{D}=0$).

In the cases with vegetation, the shear stress at the wall is given by the sum of the mean canopy drag and the viscous shear stress there, $\overline{\tau_w} = \overline{\tau_{tot}}(0) = \overline{D}(0) + \overline{\tau_1}(0)$, as all the other contributions vanish due to the no--slip and no--penetration boundary conditions. The relative magnitude of $\overline{D}(0)$ and $ \overline{\tau_1}(0)$ is likely set by the ratio between the volume occupied by stems and that occupied by stems and fluid within the canopy (i.e., the volume fraction), which for us is about $16\%$. Moving upward, away from the wall, the canopy drag $\overline{D}$ and the viscous shear $\overline{\tau_1}$ decrease, while all the other components grow due to the onset of the {mean flow} and of the turbulent fluctuations. Notably, the growth of the turbulent shear $\overline{\tau_4}$ is non--monotonous in the most rigid case, $Ca=10$, while it appears more regular at  $Ca=100$, compatibly with an increased turbulent activity within the canopy in the latter case due to the flapping motion of the stems and the deeper penetration of external fluctuations. The profiles of the viscous shear $\overline{\tau_1}$ and of the turbulent shear $\overline{\tau_4}$ differ from those observed in a full canopy due to the averaging across the vegetated and non--vegetated regions. Nevertheless, the viscous shear $\overline{\tau_1}$ always exhibits a local maximum in correspondence of the drag discontinuity at the canopy tip, where the canopy drag $ \overline{D}$ goes to zero. The peak of $\overline{\tau_1}$ is more spread in the most flexible case since the enhanced compliance of the stems to the flow yields a weaker shear--layer at their tip. The {mean flow} advection $\overline{\tau_4}$ grows until saturation is attained above the canopy, where the {mean flow} is more intense. Viscous diffusion $\overline{\tau_2}$ is negligible everywhere. Finally, all contributions decay to zero at the free-slip surface to satisfy the symmetry constraint. 

\begin{figure}
    \includegraphics[width=\textwidth]{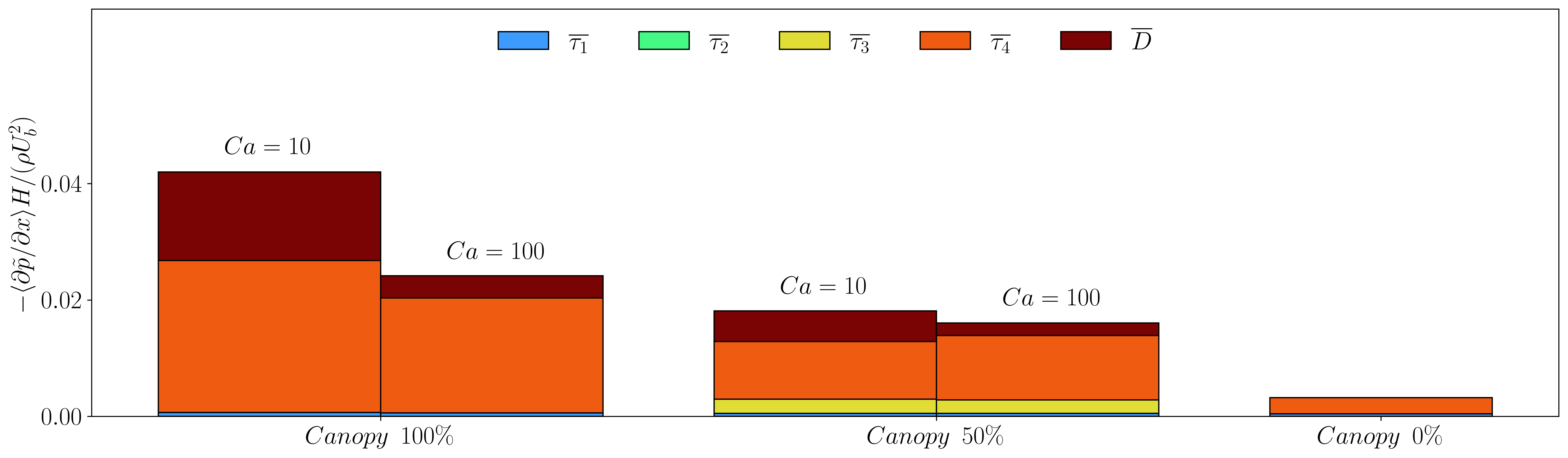}
    \caption{Wall-normal integrals of the averaged shear--balance terms computed from our simulations of a full canopy (canopy 100\%), a partially-obstructed channel (canopy 50\%), and an open channel (canopy 0\%), with the first and the last from our former investigation \citep{foggirota-etal-2024-2}. The different contributions sum to the total pressure gradient needed to sustain a fully turbulent flow at $\Rey_b=5000$ in each setup. All flow parameters are matched between each set of simulations. We also ensure the matching of the structural parameters in the vegetated cases, for which we report two different values of stem flexibility. $\overline{\tau_1}$ is the viscous shear stress, $\overline{\tau_2}$ is the viscous diffusion of streamwise momentum in the spanwise direction, $\overline{\tau_3}$ is the advection of streamwise momentum by the {mean flow} in the $y$--$z$ plane, $\overline{\tau_4}$ is the turbulent shear stress, and $\overline{D}$ is the mean canopy drag.}
    \label{fig:shearHist}
\end{figure}
A more direct quantification of the resistance opposed by the substrate to the flow is achieved integrating the averaged stress components along the wall normal direction, attaining different scalar contributions which sum to the driving pressure gradient. Those are shown in the histograms of figure~\ref{fig:shearHist} for the different cases considered. We first notice that, expectably, the highest drag is attained in the most rigid case of the full canopy, followed by the more compliant one, the two partially obstructed channels ordered by increasing $Ca$, and finally the open channel case. The viscous shear contribution $\overline{\tau_1}$ is small and essentially unchanged across all cases, while the viscous diffusion $\overline{\tau_2}$ is totally negligible in agreement with the previous plots (figure~\ref{fig:shearBalance}). The advection of streamwise momentum $\overline{\tau_3}$ accounts for part of the total drag in the partially obstructed channel cases, slightly decreasing for the most compliant stems, while the most significant contribution comes from the turbulent shear stress $\overline{\tau_4}$ in all cases. Differently from the full canopy, where the turbulent shear is depleted increasing the flexibility of the stems, this does not appear to be the case in the partially obstructed channel. There, the turbulent shear generated by the discontinuity between the vegetated and the non--vegetated regions overshadows the effect of a variation in the stem flexibility. The flexibility effect is thus limited to the canopy drag $\overline{D}$, which undergoes a decrease of more than $50\%$ increasing from $Ca=10$ to $Ca=100$ mainly due to the deflection of the stems and the consequent reduction of the frontal area of the canopy.

\subsection{Turbulent structures and events} \label{sSec:structs}

\begin{figure}
\centering
\begin{subfigure}{\linewidth}
    \centering
    \includegraphics[angle=90, width=.4\textwidth]{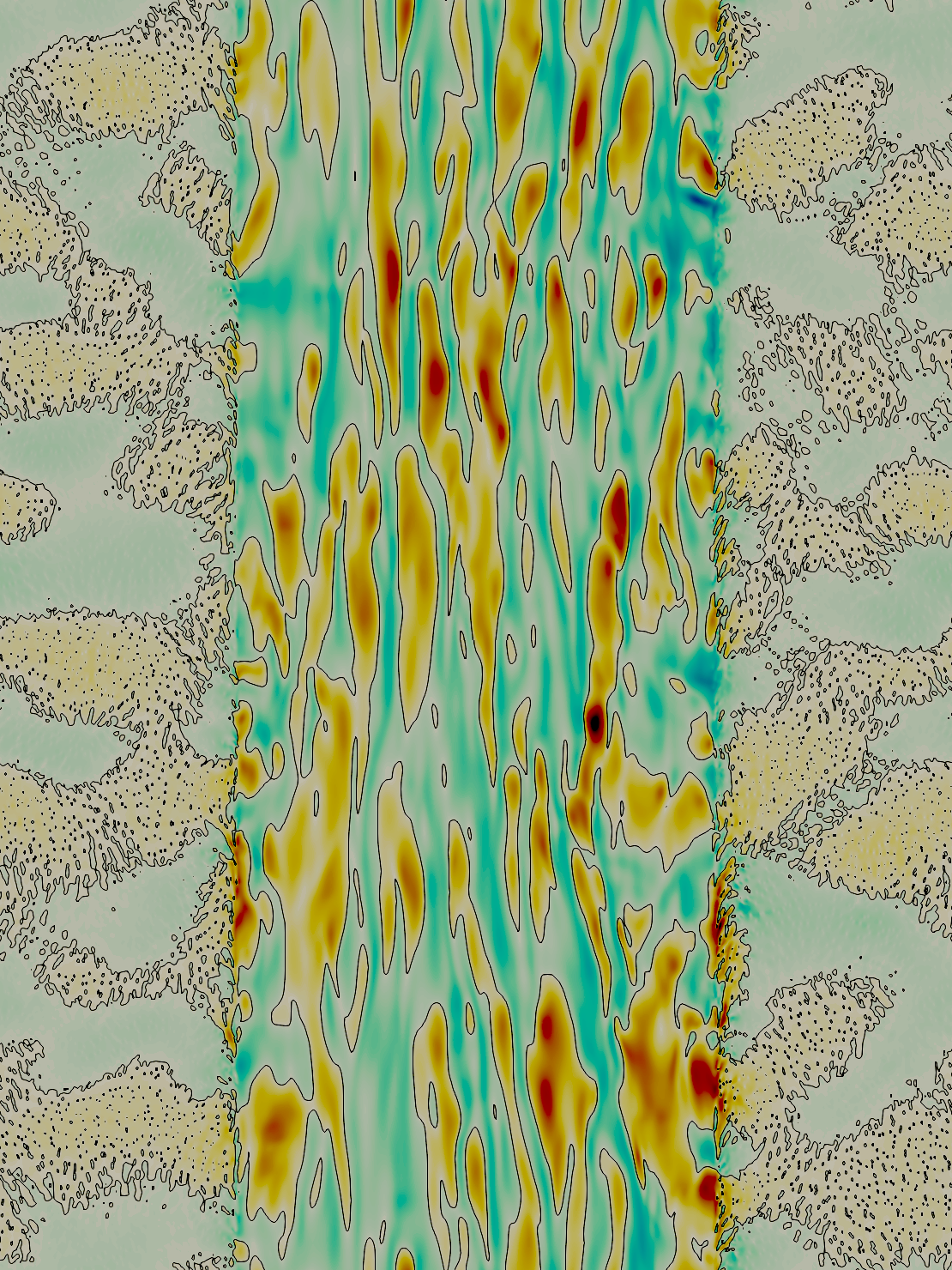}
    \hspace{1cm} 
    \includegraphics[angle=90, width=.4\textwidth]{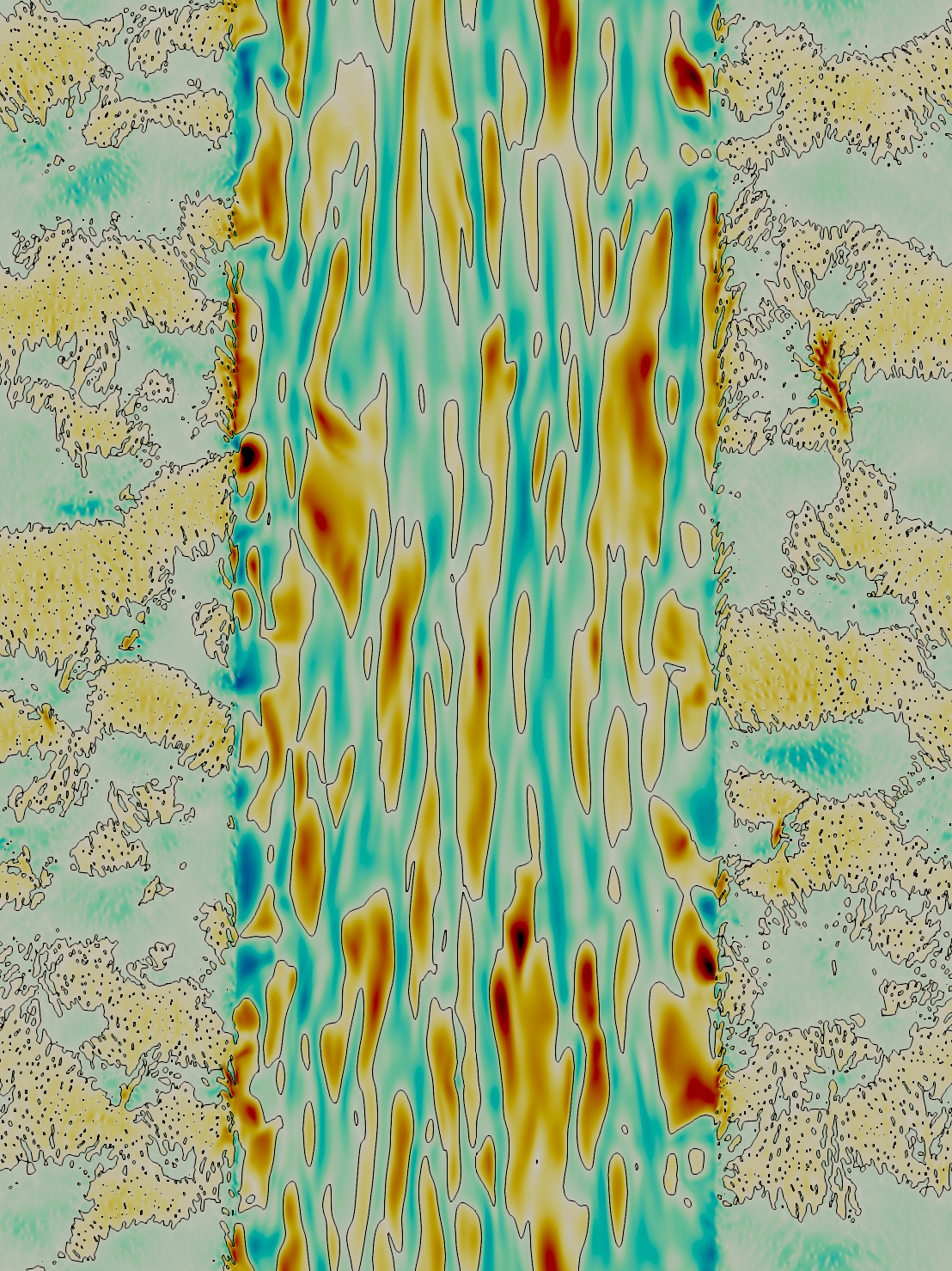}
    \put(-340,120){$Ca = 10$}
    \put(-152,120){$Ca = 100$}
    \put(-355,110){$u'$}
    \vspace{1cm} 
\end{subfigure}
\begin{subfigure}{\linewidth}
    \centering
    \includegraphics[angle=90, width=.4\textwidth]{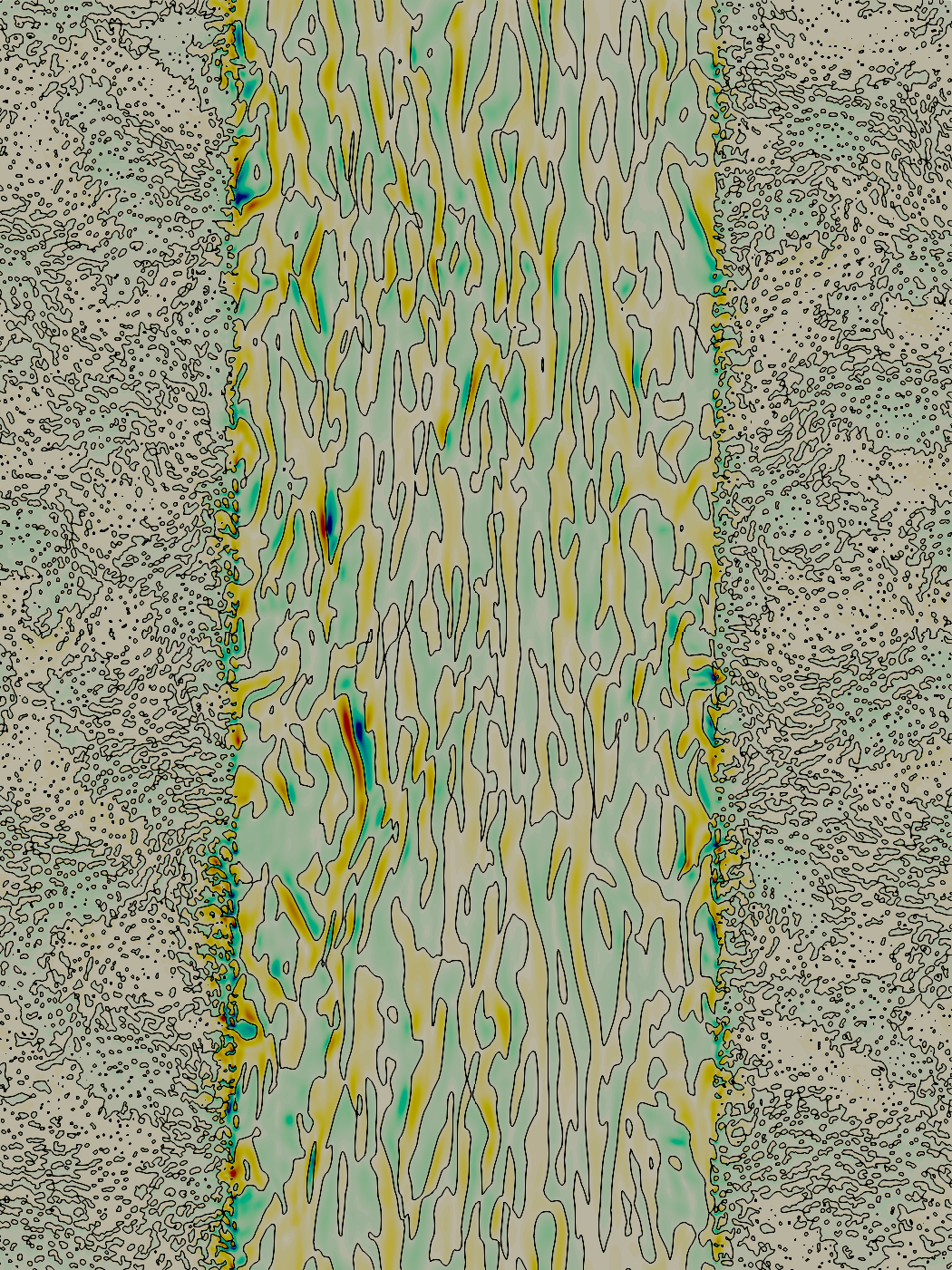}
    \hspace{1cm} 
    \includegraphics[angle=90, width=.4\textwidth]{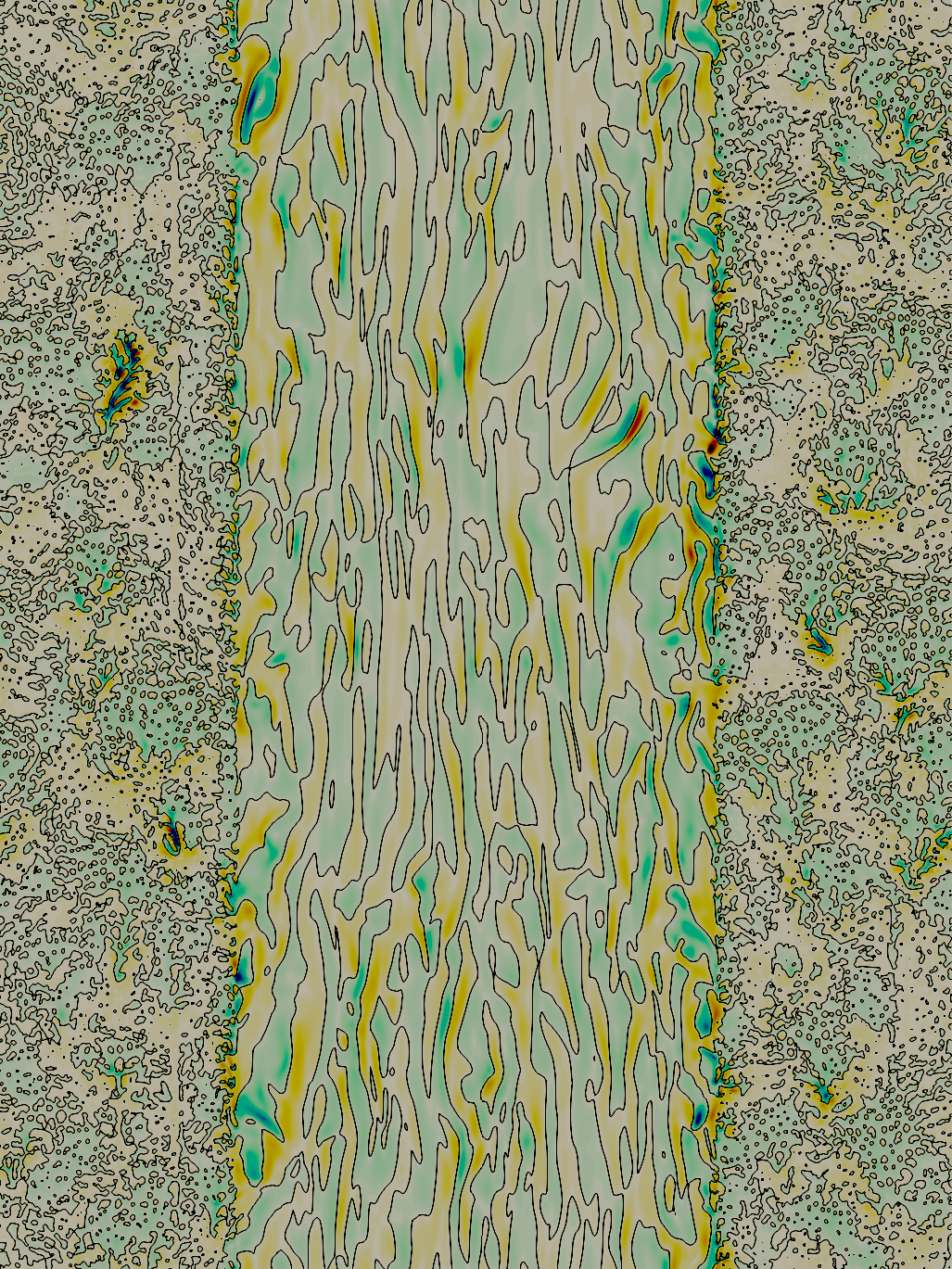}
    \put(-355,110){$v'$}
    \vspace{1cm} 
\end{subfigure}
\begin{subfigure}{\linewidth}
    \centering
    \includegraphics[angle=90, width=.4\textwidth]{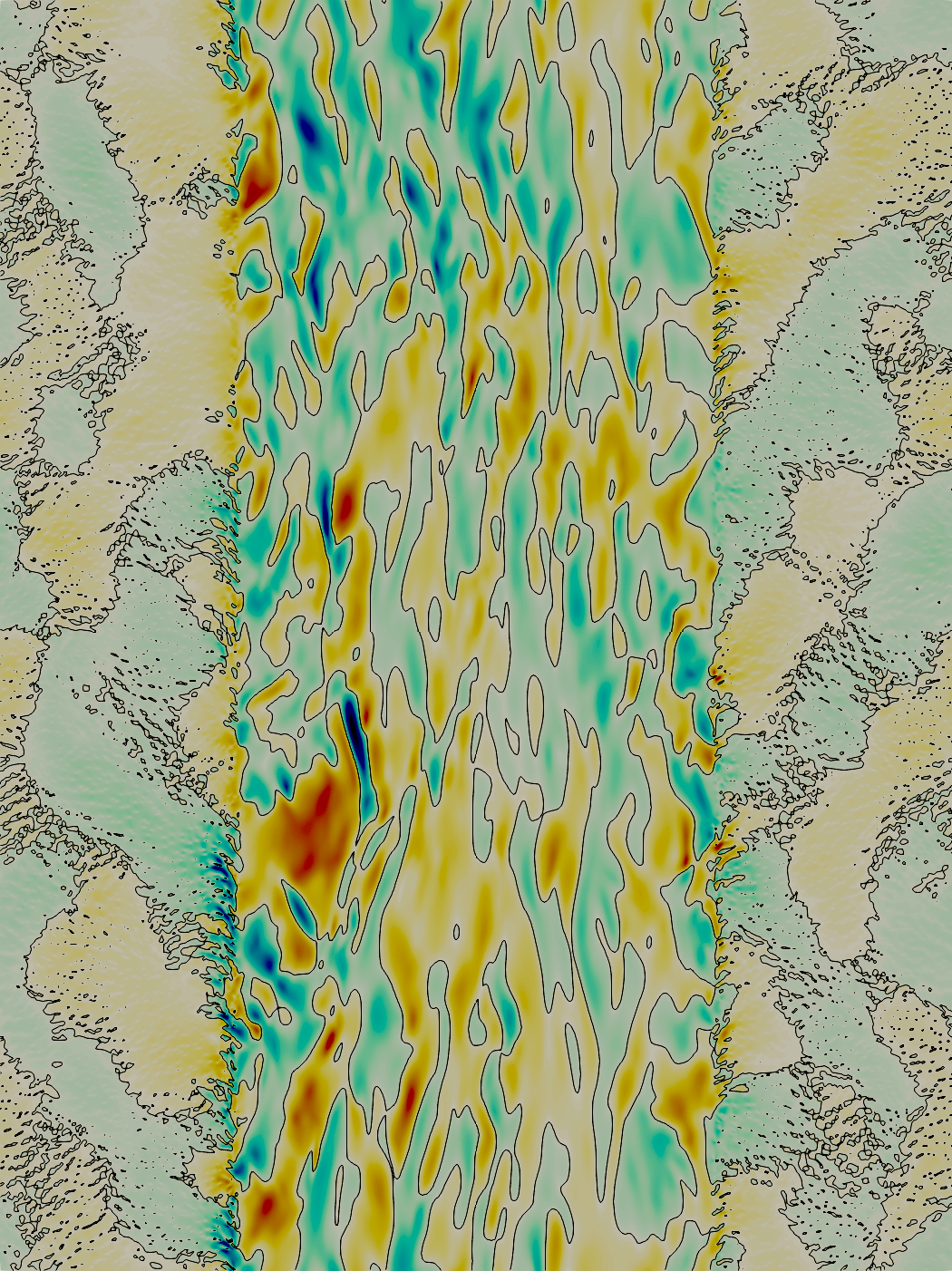}
    \hspace{1cm} 
    \includegraphics[angle=90, width=.4\textwidth]{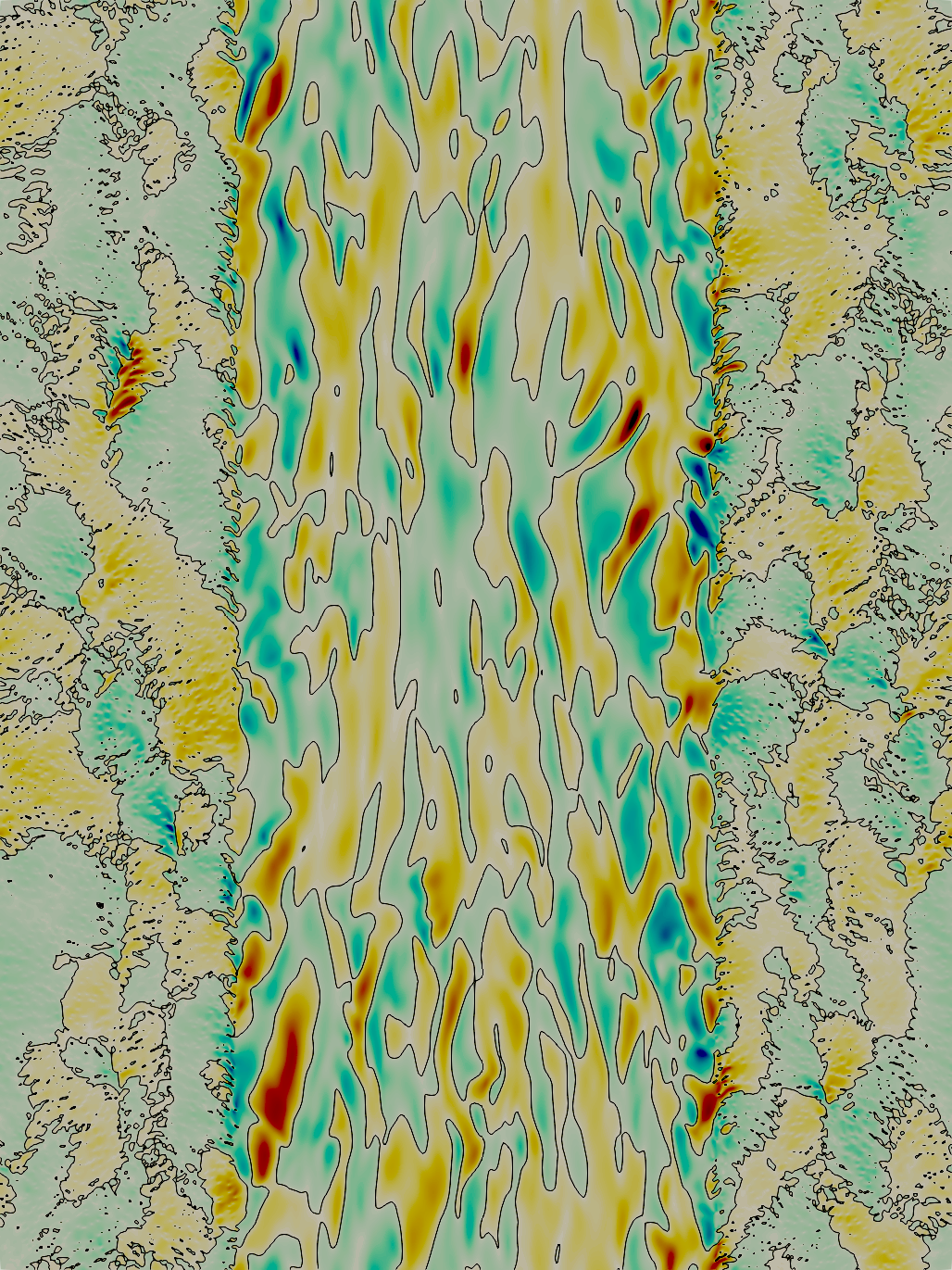}
    \put(-355,110){$w'$}
\end{subfigure}
\caption{Instantaneous flow velocity fluctuations in a wall--parallel plane at $y=0.01H$ of our partially obstructed channels, with the flow going from left to right. The left column refers to the case with the most rigid stems, the right column to that with the most compliant ones. In the first row, we show the streamwise velocity component, ranging from blue to red within $\pm 0.5 U_b$; in the second row, we show the wall--normal velocity component, ranging from blue to red within $\pm 0.2 U_b$; in the third row, we show the spanwise velocity component, ranging from blue to red within $\pm 0.5 U_b$.}
\label{fig:wallSlices}
\end{figure}

Different regions of the flow are populated by different kinds of coherent turbulent motions, interacting one with the other. We start observing those close to the bottom wall, slicing the domain with a wall--parallel plane at $y=0.01H$; the different components of the fluctuating velocity $\mathbf{u'}$ there are shown in figure~\ref{fig:wallSlices}. The streamwise component $u'$ within the vegetation gap (the central ``band" of the domain) highlights the formation of the typical low-- and high--speed streaks found in wall turbulence \citep{smith-metzler-1983}, while within the canopy we observe large regions of coherent motion extended in the spanwise direction. Their coherence is reduced increasing the flexibility of the stems, due to the disruptive effect of their flapping. A similar effect is observed for the spanwise component $w'$, where nevertheless the coherent regions within the canopy exhibit a more oblique arrangement. Comparing $u'$ and $v'$ within the canopy, we better understand their spatial coherence with the formation of local zones of flow divergence \citep{monti-etal-2020} where the vertical momentum penetrating from the canopy tip gets redistributed along the wall--parallel directions. Accordingly, the wall--normal component $v'$ within the canopy close to the wall is small and incoherent, since it gets disrupted by the stems. Only in the most flexible case we witness few intense jets of wall--normal velocity reaching the wall. This is not the case of the flow within the vegetation gap: there, all the velocity components exhibit the most intense fluctuations approaching the interface between the vegetated and non--vegetated regions, due to the enhanced turbulent activity.

\begin{figure}
\centering
\begin{subfigure}{\linewidth}
    \centering
    \includegraphics[width=.45\textwidth]{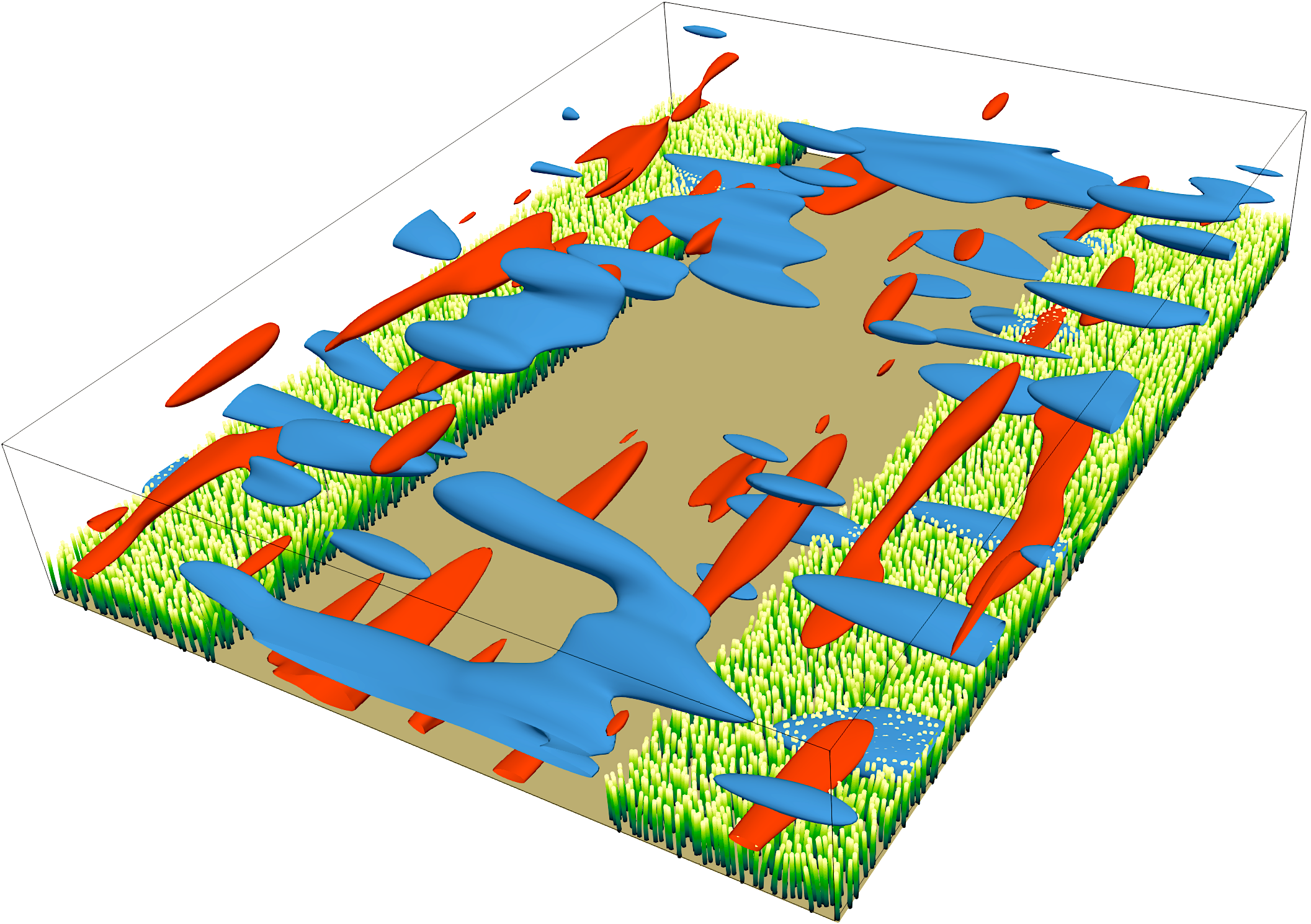}
    \hspace{.1cm} 
    \includegraphics[width=.45\textwidth]{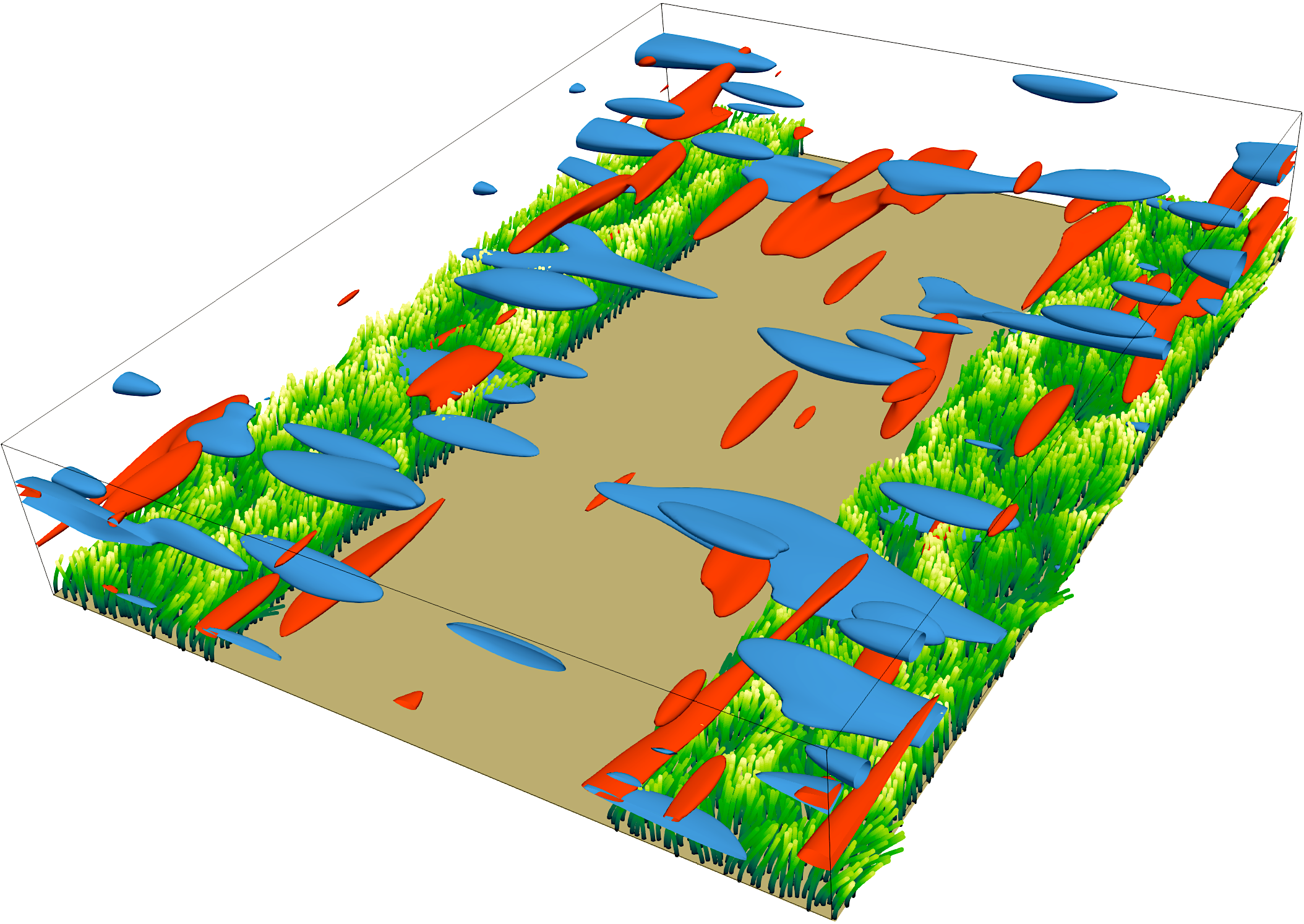}
    \put(-280,130){$Ca = 10$}
    \put(-100,130){$Ca = 100$}
    \put(-350,80){$\omega'$}
    \vspace{1cm} 
\end{subfigure}
\begin{subfigure}{\linewidth}
    \centering
    \includegraphics[width=.45\textwidth]{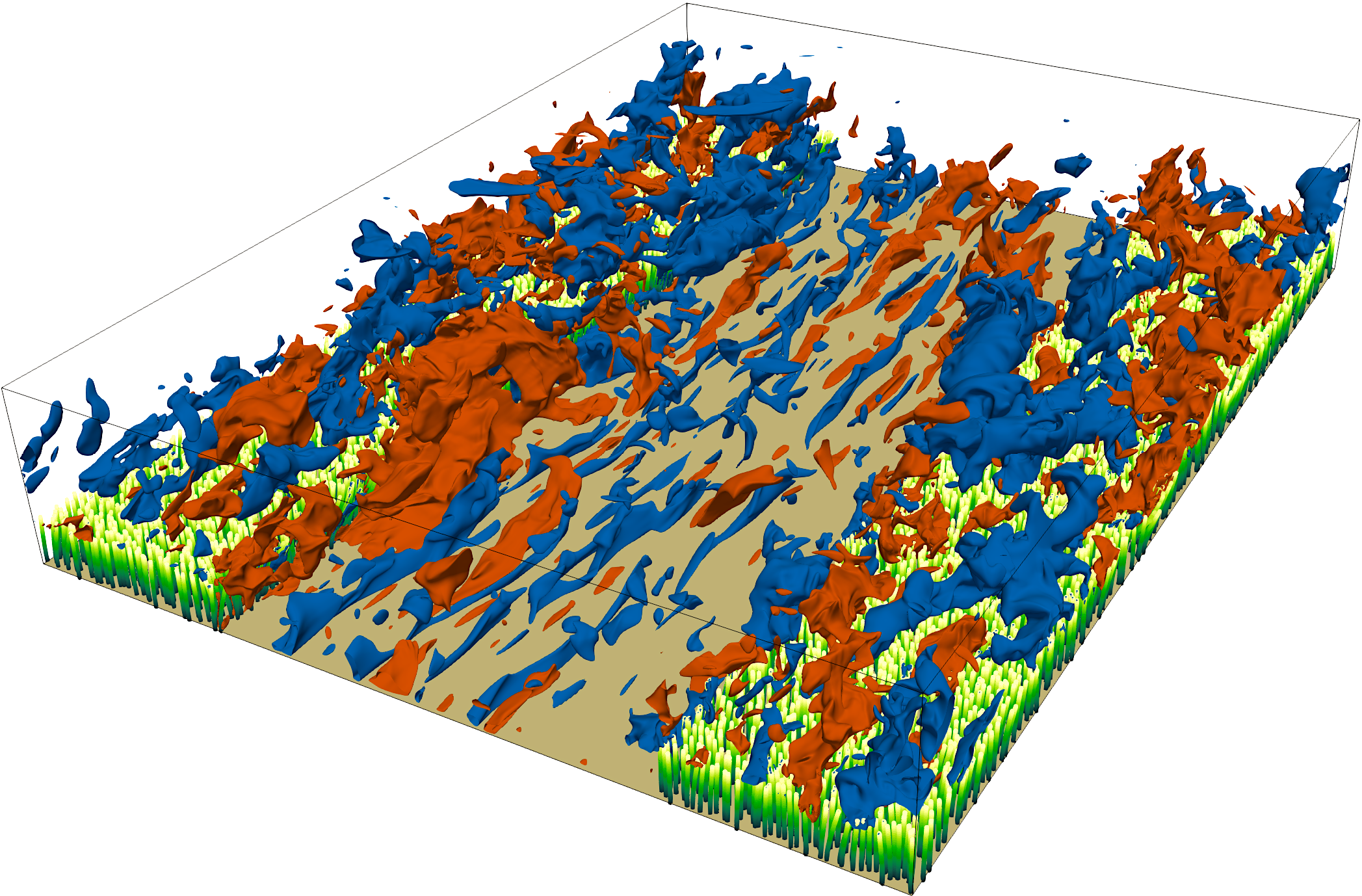}
    \hspace{.1cm} 
    \includegraphics[width=.45\textwidth]{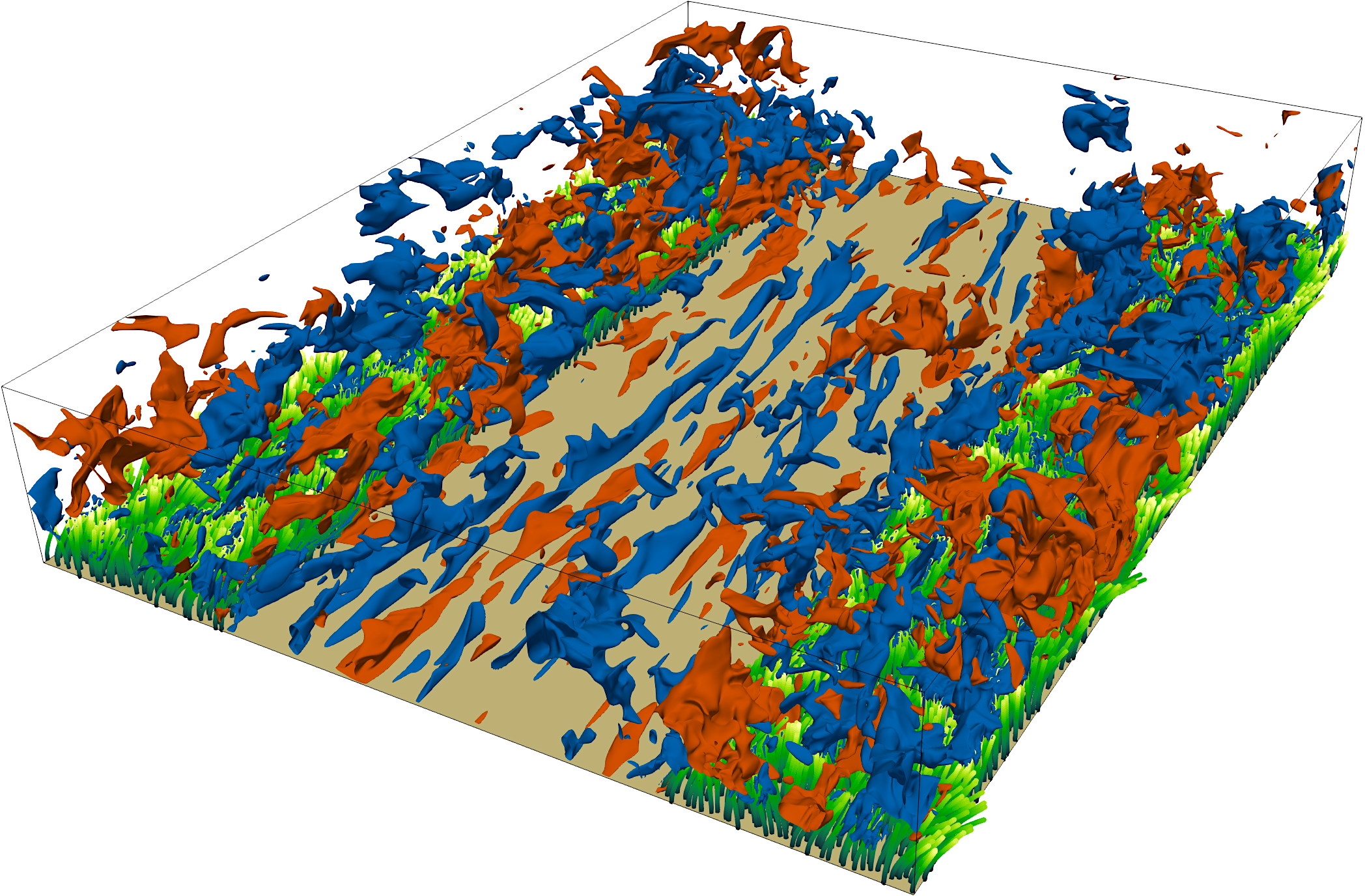}
    \put(-350,80){$u'$}
\end{subfigure}
\caption{Visualisations of the instantaneous flow field in our partially obstructed channels, with the flow going from left to right aligned with the vegetation gap. The left column refers to the case with the most rigid stems, the right column to that with the most compliant ones. In all plots, the elevation of the stems between $y=0$ and $y=0.25H$ is denoted varying their colour continuously from green to white. In the first row, we have filtered the fluctuating velocity field $\mathbf{u'}$ as discussed in the main text, and thus we show the iso--surfaces of its streamwise (red) and spanwise (blue) vorticity $\tilde{\boldsymbol{\omega'}}$ at a fixed value of $1 U_b/H$. In the second row, instead, we directly report the iso--surfaces of the streamwise velocity fluctuations $u'$ at $-0.3 U_b$ (blue) and $0.3 U_b$ (red).}
\label{fig:visl}
\end{figure}

We now investigate the turbulent structures generated above the canopy tip and in the vegetation gap. The presence of intense vortices above the vegetation (and rough/porous substrates in general) is well established and understood \citep{finnigan-2000,jimenez-etal-2001}, yet their extension towards the vegetation gap and their modulation with the stem flexibility is yet to clarify in the present case. In order to extract them, we thus deploy two different box filters: the first one, with a preferential streamwise orientation, has a cuboidal kernel made by $61\times15\times15$ grid cells along the streamwise, wall--normal, and spanwise directions, respectively. The second one, with a preferential spanwise orientation, has a cuboidal kernel made by $15\times61\times15$ grid cells. After applying the first filter to an instantaneous field of the fluctuating velocity $\mathbf{u'}$, we thus compute its streamwise vorticity $\tilde{\omega_x}$ and report its iso--surfaces in red in the top panels of figure~\ref{fig:visl}. Similarly, after applying the second filter to the same instantaneous field of the fluctuating velocity $\mathbf{u'}$, we compute its spanwise vorticity $\tilde{\omega_z}$ and report its iso--surfaces in blue in the top panels of figure~\ref{fig:visl}. Qualitatively similar results are attained varying the kernel size of the filter within a reasonable range. We are thus able to observe intense spanwise--oriented rollers spanning the flow region immediately above the canopy tip, inducing secondary streamwise--oriented vortices \citep[much like in figure~12 of][]{finnigan-2000}. In the most rigid case, the strong drag discontinuity at the canopy tip induces the formation of intense rollers, eventually able to merge across the vegetation gap and to span almost the whole domain transversally. In the most flexible case, instead, the rollers appear limited to the vegetated region and its immediate neighbourhood.

Also for the streamwise--oriented vortices we observe a more homogeneous distribution throughout the domain in the case at $Ca=100$. One of the processes generating them is in facts related to the secondary instability of the spanwise rollers \citep{finnigan-2000}, and they thus occupy similar regions. In between pairs of streamwise--oriented vortices, streak--like regions of high and low streamwise velocity $u'$ are generated. We thus visualise them in the bottom panels of figure~\ref{fig:visl} for the different values of stem rigidity we considered, denoting in blue a negative velocity fluctuation and in red a positive one. Indeed, we observe large elongated structures over the canopy, visibly responsible for the deflection of the stems below them in the most flexible case. We also appreciate an alternative visualisation of the ``typical" streaks close to the wall within the gap, with respect to that provided in figure~\ref{fig:wallSlices}. At the interface between the vegetated and non--vegetated regions, in the most rigid case, we notice the generation of larger structures likely associated to edge effects, yet we cannot confirm their presence in the most flexible case. 

\begin{figure}
\centering
    \includegraphics[width=.45\textwidth]{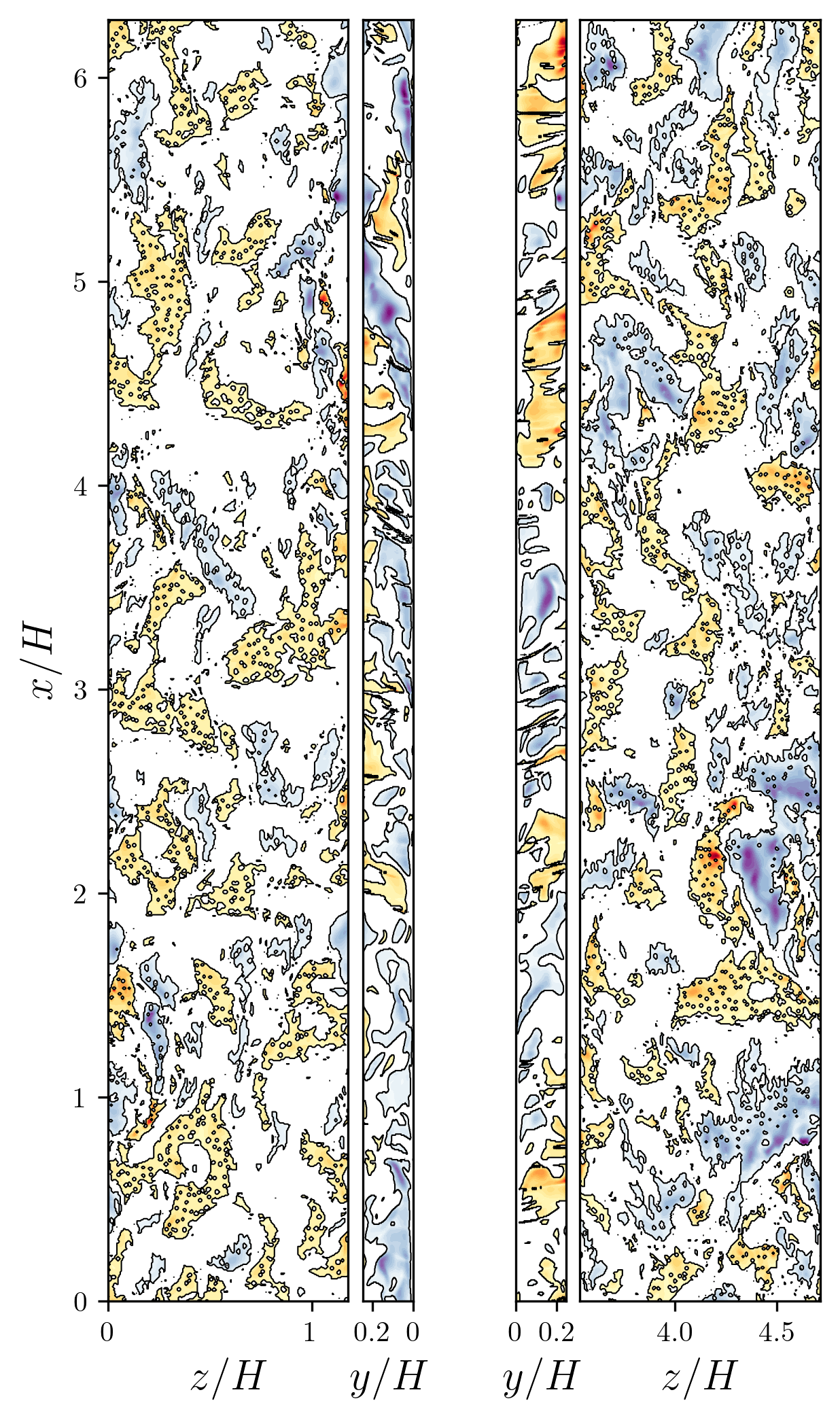}
    \hspace{1cm} 
    \includegraphics[width=.445\textwidth]{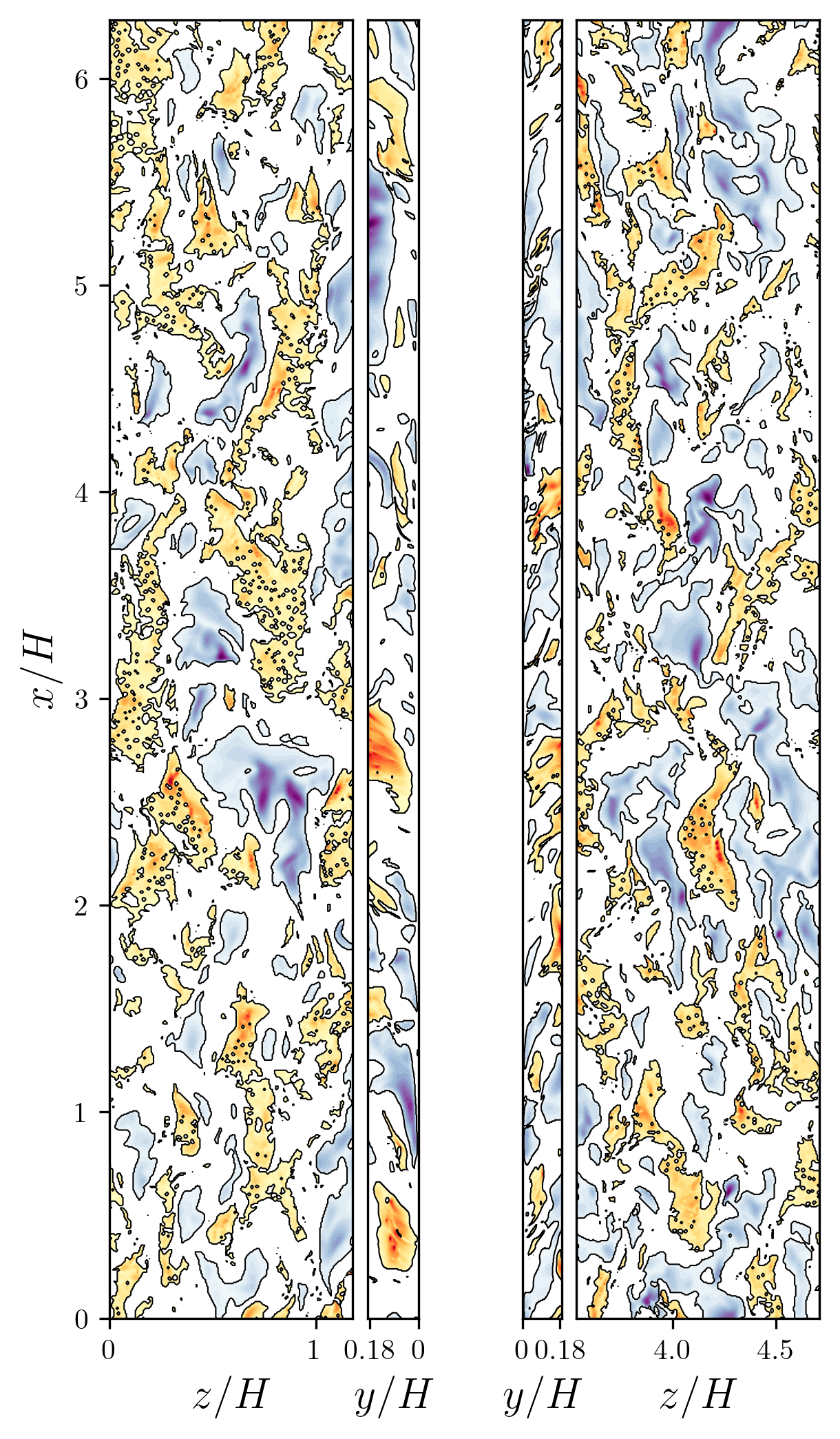}
    \put(-355,295){$Ca = 10$}
    \put(-150,295){$Ca = 100$}
\caption{Instantaneous sweep and ejection events at the average position of the canopy tip (larger rectangles) and at the interface between the vegetated and non--vegetated regions (smaller rectangles) for the the two different values of $Ca$ considered in our study. We sample the flow on wall--parallel planes (larger rectangles) and streamwise--oriented, wall--normal planes (smaller rectangles) with the {mean flow} going from bottom to top. Regions where the events are occurring are delimited with black lines, while their magnitude is quantified as $|u'v'|/U_b^2$ or $|u'w'_{out}|/U_b^2$ and visualised with a linear colour--map ranging from white to orange (ejections, in $[ 0, 0.2 ]$) or blue (sweeps, in $[ 0, 0.3 ]$).}
\label{fig:events}
\end{figure}

To better understand how momentum is exchanged between the canopy and the outer flow, we now turn our attention to the events taking place at the interface between the two (on top and on the side of the canopy). Sweep events at the average position of the canopy tip are defined as flow regions where $u'>0$ and $v'<0$, while ejections have $u'<0$ and $v'>0$ \citep{gao-shaw-paw-1989,poggi-katul-albertson-2004}. At the left interface between the vegetated and non--vegetated regions, sweeps have $u'>0$ and $w'<0$ while ejections have $u'<0$ and $w'>0$, yet at the right interface sweeps have $u'>0$ and $w'>0$ while ejections have $u'<0$ and $w'<0$. To resolve this ambiguity, we denote with $w'_{out}$ the fluctuations of the spanwise velocity going out from the canopy and use for both interfaces the definition natural to the left one. Instantaneous visualisations of the events are provided in figure~\ref{fig:events}. At the canopy tip, as already noticed in our previous work \citep{foggirota-etal-2024-2}, sweep events appear less frequents than ejections, but often have higher intensity. Increasing the flexibility of the stems, the events grow larger and more spatial coherence is attained. An analogous behaviour is found at the lateral interface between the canopy and the gap, where we notice that the most intense events take place far away from the wall, compatibly with the enhanced turbulent activity observed there (see figure~\ref{fig:normalStresses}).

\begin{figure}
\centering
    \includegraphics[width=.49\textwidth]{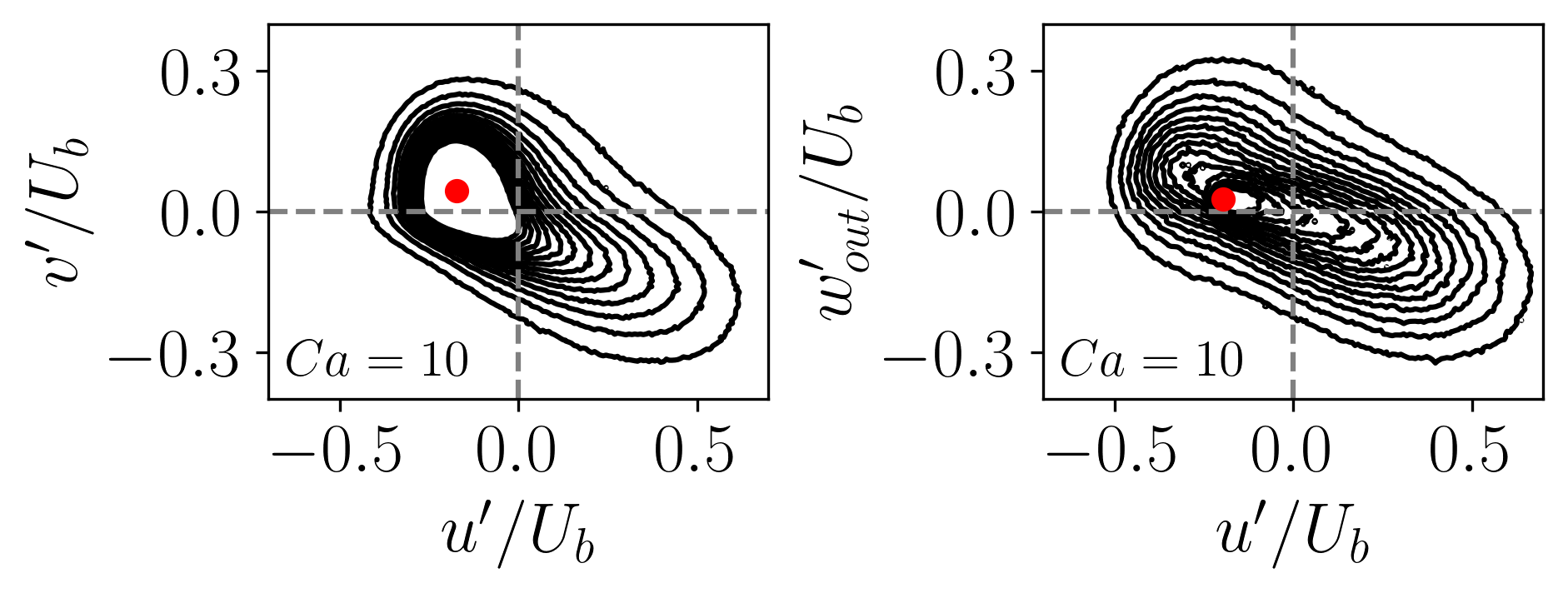}
    \includegraphics[width=.49\textwidth]{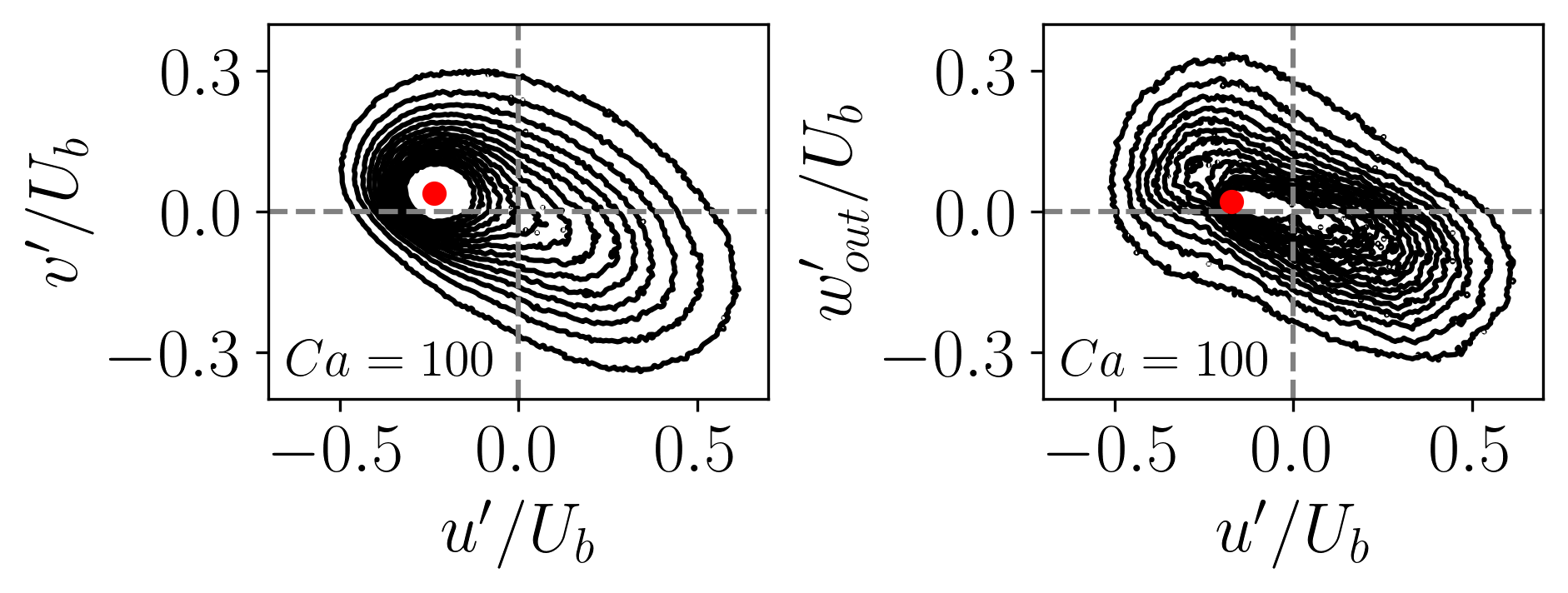}
    \put(-348,75){$Ca = 10$}
    \put(-158,75){$Ca = 100$}
\caption{Isolines of the J-PDFs (normalised to a unitary integral over the domain) associated to the fluctuations of the streamwise velocity component and to the velocity component going out from the canopy at the planes selected in figure~\ref{fig:events}. We consider the mean position of the canopy tip ($u'v'$ panels) and at the interface between the vegetated and non--vegetated regions ($u'w'_{out}$ panels), for the the two different values of $Ca$ employed in our study. Levels are evenly distributed between 0.4 and 6 with 0.4 increments, while the locations of the peaks are denoted by red dots.}
\label{fig:JPDFs}
\end{figure}

To provide a more quantitative description of the events, we now observe the joint probability--density--functions (J-PDFs) of the fluctuating streamwise velocity component and of the fluctuating velocity component exiting the canopy, within the planes selected above. Those are reported in figure~\ref{fig:JPDFs}. Ejections are the most frequent events at the tip of the canopy, while sweeps are rarer but stronger: the canopy opposes little hindrance to the fluid being lifted from behind the rollers spanning its tip, while only the most intense downwellings in front of the rollers are able to penetrate between the stems. This holds true for both the values of $Ca$ we considered, yet in the most flexible case the J-PDF exhibits a more oval shape as less streamwise momentum is lost by the fluid upon impinging on the stems. Also at the interface between the vegetated and non--vegetated regions ejections appear slightly more frequent than sweep events, with the latter reaching higher intensities. Yet the difference is significantly less sharp, suggesting that wall--normal Kelvin--Helmholtz like rollers are unlikely to drive the momentum exchange there.

\subsection{Stem reconfiguration and dynamics} \label{sSec:stems}

Investigating the dynamics of the flow, we have observed how that is affected by a variation in the stem flexibility, understanding the important role played by the reconfiguration and dynamics of the stems. Here we aim at further characterising these two aspects from the perspective of the structure. 

\begin{figure}
\centering
    \includegraphics[width=1\textwidth]{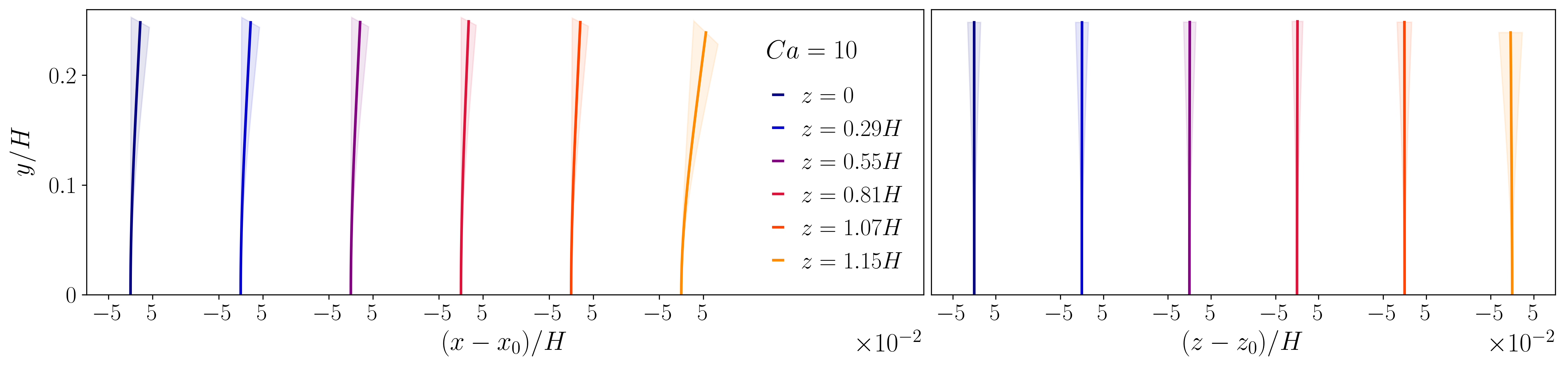}
    \includegraphics[width=1\textwidth]{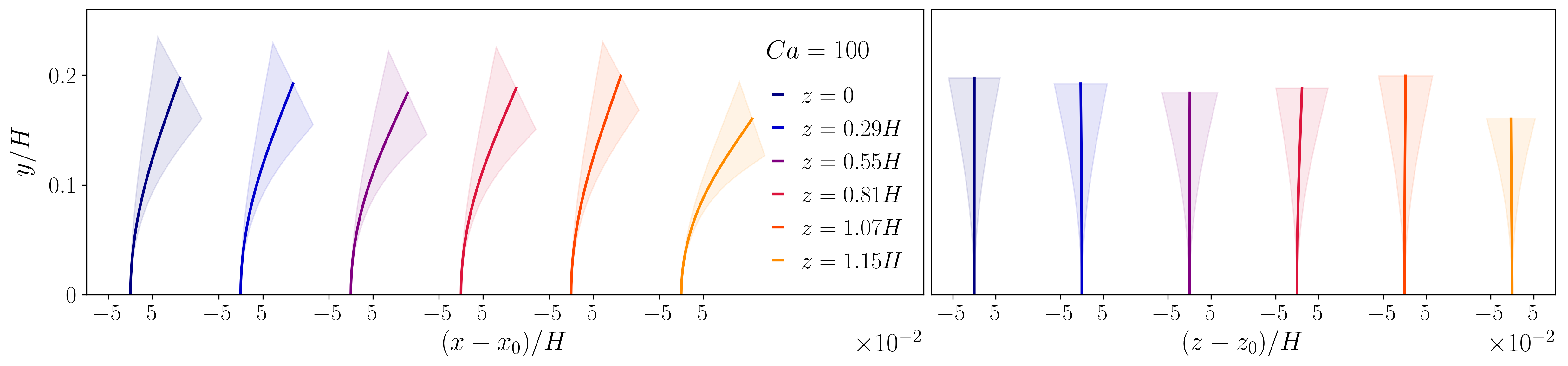}
\caption{Streamwise (left panel) and spanwise (right panels) deflection of the stems for different spanwise locations; $z=0$ lays at the middle of the vegetated region, while $z=1.15H$ lays at its margin. Results for the most rigid stems are reported in the top panels, while the bottom panels refer to the most flexible ones. The stems oscillate about their time-averaged configuration, with the shaded regions denoting the root mean square of their displacement.}
\label{fig:stemDeflection}
\end{figure}

The reconfiguration of flexible structures exposed to an incoming flow is a widely investigated topic due to its capability of reducing the drag experienced by the structure \citep{alben-shelley-zhang-2002,luhar-nepf-2011,gosselin-2019} and to modify the {mean flow} properties setting an ``implicit" boundary condition dependent on the flow itself \citep{akita-okabayashi-takeuchi-2024}. Reconfiguration is typically studied in relatively simple setups where the incoming flow, either laminar or turbulent, has no significant transversal components. Instead, here, we investigate how the structure deflects in response to the fully bi-dimensional {mean flow} established in our setup. The deflection of the stems, reported in figure~\ref{fig:stemDeflection}, is predictably more pronounced in the case at $Ca=100$, where they undergo a finite reconfiguration in the streamwise direction. Moderate reconfiguration along the streamwise direction can also be appreciated for the most rigid stems, while in no case we observe a significant deflection in the spanwise direction. The spatial pattern with which the stems bend forward, on average, is a consequence of the {mean flow} they are exposed to, and in particular of the canopy--edge vortex. Consequently, the most upright stems (located approximatively at $z=1.07H$) are found in the middle of the upwelling described in \S\ref{sSec:meanFlow}, while those located deeper in the canopy bend more, under the influence of the descending fluid. Furthermore, also the stems located closer to the interface between the vegetated and non-vegetated regions achieve a higher deflection, bending under the high-momentum fluid coming from the gap. Once deflected, the stems oscillate about their time-averaged configuration; we thus report the root mean square of their displacement with the shaded regions in figure~\ref{fig:stemDeflection}. Noticeably, in the case at $Ca=10$, oscillations of wider amplitude are achieved next to the gap, marking the intense turbulent activity associated with the canopy--edge vortex. This is not the case for the most compliant stems, which all exhibit wider fluctuations of comparable amplitude.

\begin{figure}
    \includegraphics[width=\textwidth]{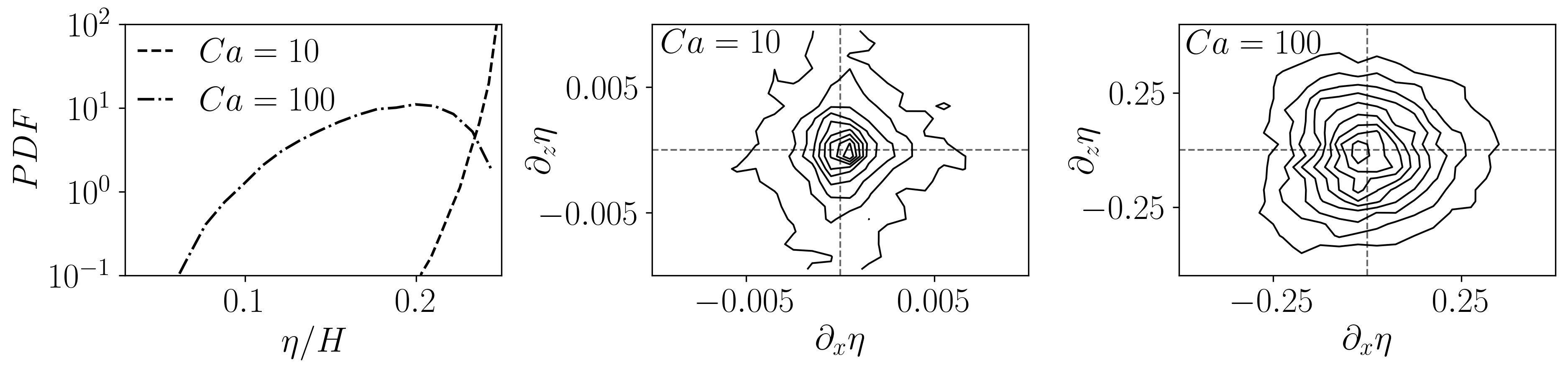}
    \caption{{Probability density functions of the canopy envelope elevation $\eta(x,y)$ for the two values of $Ca$ considered in our study (left panel). In the middle and right panels, for the same values of $Ca$, we show the joint probability density functions of the envelope gradients along the streamwise and spanwise directions.}}
    \label{fig:envelopesStats}
\end{figure}

{
Our code tracks the position of each stem tip over time, allowing us to reconstruct the full envelope of the deflected canopy tips at each time instant through a binning procedure.
We thus characterise the canopy envelopes through the probability density functions (PDFs) of their elevation $\eta(x,y)$, averaging the PDFs across all time instants. As shown in the left panel of figure \ref{fig:envelopesStats}, for $Ca=10$ the PDF exhibits a monotonous trend, quickly growing to its maximum at the value corresponding to undeflected stem ($\eta=0.25$). For $Ca=100$, instead, the PDF appears significantly broader and centred around $\eta\approx0.2$, confirming not only the increased deflection of the stems due to their higher compliance, but also their enhanced swaying motion. 
Next, in the centre and right panels of figure \ref{fig:envelopesStats}, we report the joint PDFs of the elevation gradients. The spatial derivatives of $\eta(x,y)$ have significantly smaller magnitudes at $Ca=10$ than at $Ca=100$, consistent with the reduced motion of the stems in the former case. 
Yet, both plots highlight how the spanwise gradient $\partial_z \eta$ attains larger values than the streamwise one $\partial_x \eta$, suggesting intense variations in the stem tips configuration along the transverse direction. While, in fact, only the turbulent fluctuations are responsible for rippling the canopy envelope along $x$, both the turbulent fluctuations and the {mean flow} inhomogeneity are active along $y$.} 

\begin{figure}
\centering
    \includegraphics[width=1\textwidth]{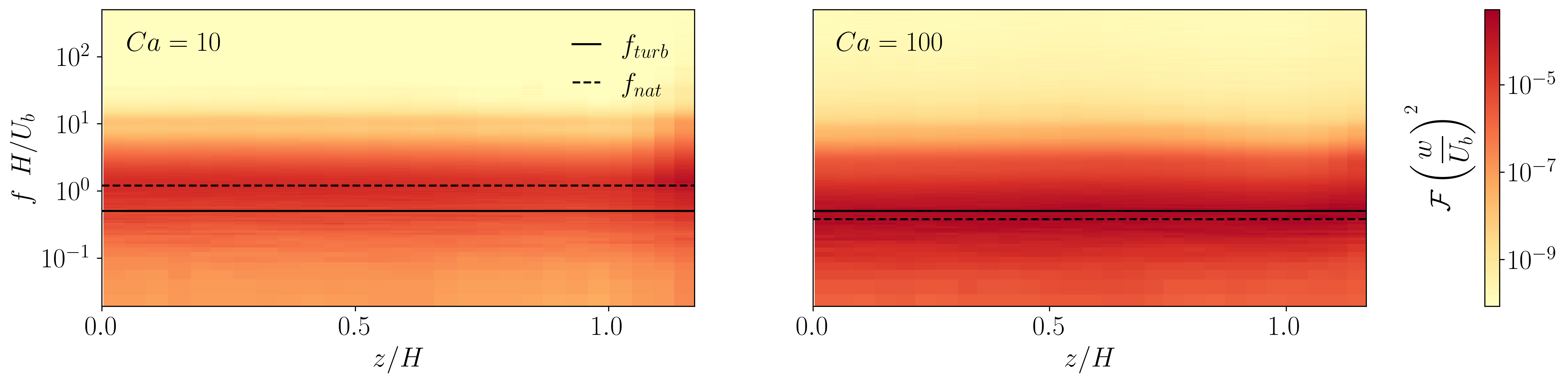}
\caption{Magnitude of the temporal response of the stems at different spanwise locations, for the most rigid (left) and most compliant (right) case. The response is computed as the squared Fourier transform of the spanwise stem--tip velocity, averaged across stems at the same spanwise location. Black lines aid the identification of the dynamical regime at which the response is occurring: $f_{nat}$ is defined as in \S\ref{sec:methods}, while $f_{turb}\approx0.5U_b/H$ following previous works \citep{foggirota-etal-2024}.}
\label{fig:stemResponse}
\end{figure}

{To link the topological features of the canopy envelope with the motion of the stems, we further probe their dynamic response in the frequency domain.}
We focus on the spanwise velocity of the stem tips as it is the Lagrangian velocity component less affected by the inextensibility of the stems and by the presence of the bottom wall. Sampling its signal for all stems over about $50 H/U_b$, we thus compute its temporal Fourier transform and average along the streamwise direction over all stems within tiles (see \S\ref{sec:setup}) occupying the same spanwise location, symmetrically across the gap. The outcome, $\mathcal{F}(w/U_b)^2$ as reported in figure~\ref{fig:stemResponse}, is a map of the response intensity at each frequency, as a function of the spanwise location of the stems. To understand the regime of oscillations of the stems, that being either their natural dynamics or a compliant response to the turbulent flow \citep{foggirota-etal-2024}, we overlay to the plots a dashed line corresponding to the structural natural frequency $f_{nat}$ (as defined in \S\ref{sec:methods}) and a continuous line corresponding to the turbulent frequency of the largest structures in the flow, $f_{turb}\approx0.5U_b/H$ \citep{foggirota-etal-2024}. From the plots, it thus appears clear that at $Ca=10$ the stem dynamics is dominated by their natural response, while the case at $Ca=100$ is close to the crossover where $f_{nat}\approx f_{turb}$. The more rigid stems oscillate with their natural dynamics and exhibit a wider range of excitation closer to the canopy-edge vortex ($z/H\gtrsim1$), where the turbulent forcing is more broad-banded. This latter observation appears consistent with the enhanced amplitude of the stem oscillations observed there, as in figure~\ref{fig:stemDeflection}. The more flexible stems, instead, are significantly more deflected and thus hindered in their natural dynamics. They therefore approach a regime where their motion is fully dominated by the turbulent forcing, and oscillate at a broader range of frequencies centred about $f_{turb}$. {The more flexible stems consequently explore a broader range of elevations, as denoted by their PDF in figure \ref{fig:envelopesStats}, and yield a more rippled envelope.}

%% file: conclusions.tex
\section{Conclusions}
\label{sec:conclusions}

In this work, we have performed and analysed DNSs of a partially--obstructed turbulent channel flow (figure~\ref{fig:setup}) along the streamwise edge of a vegetation canopy for different values of its rigidity. The turbulent flow, characterised by a bulk Reynolds number $\Rey_b=5000$, takes place between and above the flexible canopy stems, with $Ca\in\{ 10, 100\}$, as well as in the non-vegetated gap on their side.

Due to the spanwise inhomogeneity of the setup, we witness the onset of a fully bi--dimensional {mean flow} in the spanwise/wall--normal plane, characterised by the formation of an intense recirculation region aligned to the canopy edge (figure~\ref{fig:meanFlow}). While the cellular structure of the flow is reminiscent of that observed in duct flows over porous media \citep{samanta-etal-2015,suga-okazaki-kuwata-2020}, the formation of a streamwise vortex at the spanwise location of the drag discontinuity is typical of vegetation sub-layers with a finite spanwise extent \citep{moltchanov-etal-2015,song-etal-2024}. The \textit{canopy--edge vortex} draws high--momentum fluid from the gap into the side of the canopy, ejecting low--momentum fluid from the canopy tip in an upwelling close to the canopy edge. This mechanism, in agreement with former observations for vegetated \citep{yan-etal-2016} and rough beds \citep{stroh-etal-2020}, proves sensitive to the spanwise extent of the canopy and appears inverted in cases with higher confinement \citep{unigarovillota-etal-2023}. 

Despite its appreciable effect on the {mean flow}, the canopy-edge vortex is a fully turbulent phenomenon characterised by intense fluctuations of all the velocity components (figures~\ref{fig:normalStresses}, \ref{fig:shearStresses}). The extra--diagonal terms of the Reynolds stress tensor involving the wall--normal velocity component peak in its correspondence, and the shear stress balance is altered compared to the case of a homogeneous canopy. In facts, while the turbulent structures above the canopy are significantly altered by a variation in the flexibility of the stems (figure~\ref{fig:visl}), the canopy--edge vortex is not particularly affected. Consequently, while for a homogeneous canopy the turbulent contribution to the total drag (figure~\ref{fig:shearHist}) exhibits a pronounced dependance from the flexibility of the stems as most of the fluctuations are generated close to the canopy tip, for the partially--obstructed channel the turbulent contribution does not vary significantly with the stem flexibility, as it mostly comes from the canopy--edge vortex. Despite the unavoidable generation of a drag discontinuity at the edge between the canopy and the gap, we do not witness the onset of horizontal coherent vortices there \citep{yan-etal-2022-1} associated to a Kelvin--Helmholtz like instability, as we find over the canopy tip. Consistently, the joint probability density functions of the velocity fluctuations (figure~\ref{fig:JPDFs}) appear different at the tip and at the edge of the canopy, suggesting that momentum transfer there is regulated by the canopy--edge vortex instead. As a consequence of the intense turbulent activity along the canopy edge, the stems located there exhibit more intense oscillations on a broader frequency range compared to those located deeper inside the canopy (figures~\ref{fig:stemDeflection}, \ref{fig:stemResponse}), in the most rigid case. 

Further investigations should better elucidate the dependence of the results from the spanwise extent of the vegetated and non--vegetated regions, and address how the formation of horizontal coherent vortices at the canopy edge is affected by the choice of the flow and structural parameters. In sight of potential practical applications, it would also be of interest to understand how plants of more complex morphology alter the scenario described here.

%% file: validation.tex
\section{Validation}
\label{app:validation}

\begin{figure}
\centering
\includegraphics[width=.95\textwidth]{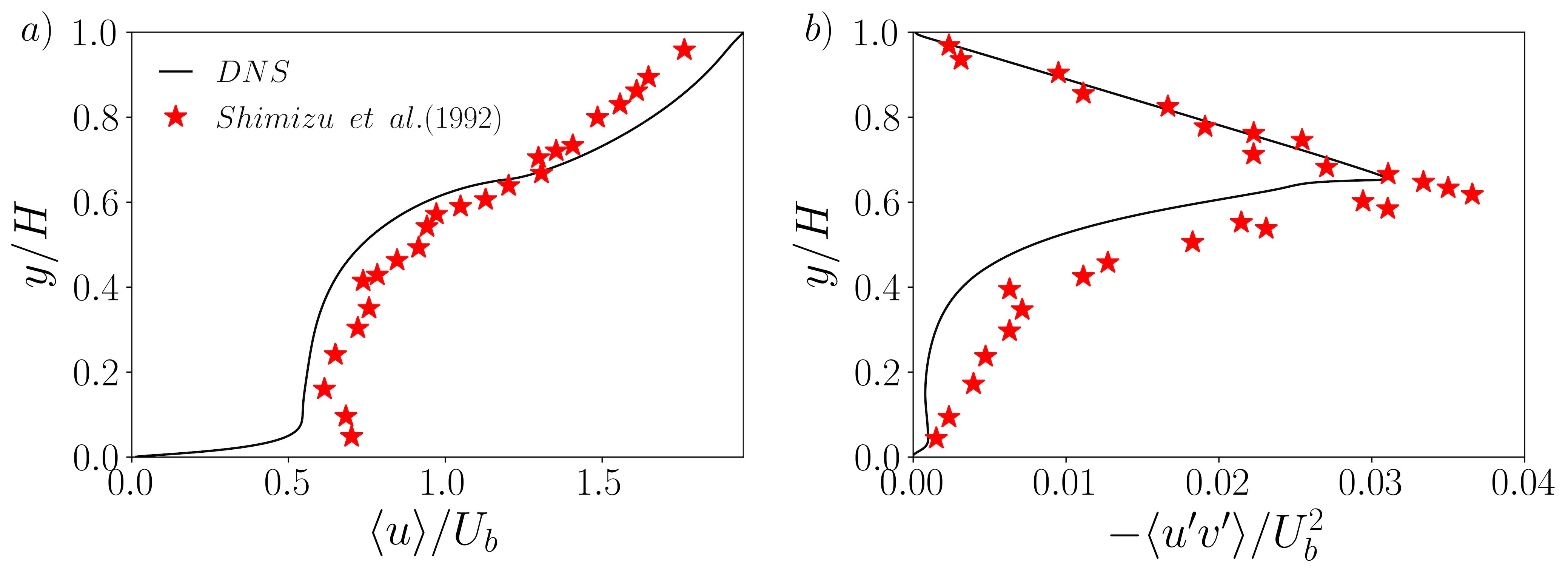}
\caption{Mean streamwise velocity profile ($a$) and Reynolds shear stress ($b$) in and above a dense and rigid canopy, from \cite{monti-olivieri-rosti-2023}. Red stars denote the experimental measurements of \cite{shimizu-etal-1992}, while black lines are the outcome of a DNS matching the experimental parameters, performed with our code.} 
\label{fig:expProf}
\end{figure}

The configuration of our full canopy simulations is in agreement with the experiments of \cite{shimizu-etal-1992}, where a rigid and dense canopy with height $h=0.65H$ is exposed to a turbulent flow at $Re_b=7070$. Simulations of the flow within and above a rigid canopy exactly matching these parameters, performed with the same code we used \citep{monti-olivieri-rosti-2023}, allow us to compare the computed {mean flow} profile and Reynolds shear stress with the experimental data. This comparison, {reported in figure \ref{fig:expProf}}, confirms that our code reliably captures the back-reaction of the structure on the fluid.

